\documentclass[aps,preprint,onecolumn,nofootinbib]{revtex4}
\usepackage{tipa,mathrsfs}
\usepackage{bm}
\usepackage{mathrsfs}
\usepackage{amssymb}
\usepackage{amsmath}
\usepackage{amsfonts,dsfont}
\usepackage{color}
\usepackage{ulem}

\usepackage[]{graphicx}
\begin{document}
\newcommand{\half}{\frac{1}{2}}
\title{Whence Nonlocality? Removing spooky action at a distance from the de Broglie Bohm  pilot-wave theory using a time-symmetric version of de Broglie double solution}
\author{Aur\'elien Drezet$^{1}$}
\address{$^1$Univ. Grenoble Alpes, CNRS, Grenoble INP, Institut Neel, F-38000 Grenoble, France}

\email{aurelien.drezet@neel.cnrs.fr}
\begin{abstract}
In this work, we review and extend a version of the old attempt made by Louis  de broglie for interpreting quantum mechanics in realistic terms, namely the double solution. In this theory quantum particles are localized waves, i.e, solitons, that are solutions of relativistic nonlinear field equations. The theory that we present here is the natural extension of this old work and relies on a strong time-symmetry requiring the presence of advanced and retarded waves converging on particles. Using this method, we are able to justify wave-particle duality and to explain the violations of Bell's inequalities. Moreover, the theory recovers the predictions of the pilot-wave theory of de Borglie and Bohm, often known as Bohmian mechanics. As a direct consequence, we reinterpret the nonlocal action at a distance presents in the pilot-wave  theory. In the double solution developed here there is fundamentally no action at a distance but the theory requires a form of superdeterminism driven by time-symmetry.   
\end{abstract}

\maketitle
\section{Preambule}\label{sec1}
\indent One hundred years ago  the genious of Louis de Broglie  gave birth to quantum wave mechanics that we celebrate in this special issue of the journal Symmetry. The consequences of de Broglie insights \cite{debroglie1923a,debroglie1923b,debroglie1923c,debroglie1924,debroglie1925th}, particularly his understanding in 1923 of the great generality of wave-particle duality that applies to every  kinds of material particles (i.e., not only limited to particles of light as postulated  by Einstein already in 1905) paved the way for  modern quantum mechanics  and all its physical and technological implications. Moreover, de Broglie intuition was based on a realistic and deterministic picture in which particles follow trajectories in space-time. His view contrasts, and even conflicts, with the usual description of quantum mechanics, associated with the Bohr-Heisenberg `Copenhagen interpretation' in which such a spatio-temporal and causal representation of the world is abandonned and considered as impossible.\\
\indent   It is well known that de Broglie proposed a realistic approach of quantum mechanics, namely the pilot-wave theory, that he defended in 1927 at the fifth Solvay conference \cite{debroglie1927,Valentini}. It is also  well known that he soon abandonned it because he felt his theory too paradoxical and too preliminary \cite{debroglie1930}. He thus accepted and publically advocated for 20  years the Copenhagen interpretation and came back to his realistic approach only in 1951-2 after David Bohm rediscovered a version of the pilot-wave theory (the theory is known as the de Broglie Bohm (dBB) theory \cite{Bohm1952,Hiley}). What is much less known, and  rarely discussed, is that de Broglie actually hoped to develop a different approach, namely the double solution (DS) theory presented in details in a publication of 1927 \cite{debroglie1927} and defended later in the 1950's \cite{debroglie1956}. He considered DS  as much closer to Einstein's perspective where particles are identified with localized solutions of nonlinear field equations (i.e., like in the general theory of gravitation or in the nonlinear electrodynamical theory of Mie, Born and Infeld\cite{Rosen,Mie,Born}). For de Broglie, like for Einstein,  the particle must be a concentrated amount of energy, i.e., a `soliton' or solitary wave (what Einstein called a `bunch-like' field),  moving in space-time and following a main trajectory.  De Broglie intuition, already discussed in his PhD thesis \cite{debroglie1925th}, was to assume that a particle is an oscillating soliton or pulse acting as a kind of quantum clock moving in space-time.  The relativistic properties of this clock connected with those of the soliton field solution of the nonlinear wave equation were expected to account for all the  observed quantum phenomena, e.g., including wave-particle dualism, quantum entanglement and spin properties. In particular, the central part of the soliton (hereafter the `core') was assumed to be guided by a much weaker propagating field oscillating in phase with the core of the soliton.  Originally, and partly inspired by an older proposal by Einstein where the photon was defined as a moving singularity of the electromagnetic field \cite{Einstein1909}, the DS theory considered a singular field diverging at the position of the point-like particle \cite{debroglie1927}). De Broglie expected that the trajectories predicted by the DS approach  would be equivalent to those given by the dBB pilot-wave theory which is known to reproduce standard quantum mechanics (at least in the non-relativistic limit).\\
\indent  Interestingly, a preliminary version of the DS theory using a scalar wave equation was already proposed by de Broglie in 1925 \cite{debroglie1925,debroglie1925b}, i.e., before Schr\"{o}dinger actually developped his famous wave equation. However, this older theory of de Broglie has been curiously forgotten by historians and physicists (even by de Broglie himself who never mentionned it again). When de Broglie came back to the DS approach in the 1950's and 1960's it was the 1927 version that was considered and further developed not his early  work of 1925! However, we believe that the earlier version contains a deeper and forgotten truth which must be exploited. It is the aim of the present article to pay a tribute to the remarkable intuitions of de Broglie concerning his histortical DS approach of 1925.  Here, for celebrating de Broglie insights we review and partially extend a recent proposal we did concerning this old and forgotten DS theory \cite{Drezet2023}. This new developement of de Broglie ideas constitutes, we believe,  the logical and natural completion of his early 1925 work. In fact, one of the central idea of the original double solution approach was to accept a strong time-symmetry of the fundamental field constituting the soliton \cite{debroglie1925,debroglie1925b}.   This strong time-symmetry is actually similar to the one later developped by Wheeler and Feynman \cite{Wheeler} in their famous electrodynamics namely the `absorber theory' that involves a half sum of retarded and advanced fields. De Broglie realized already in 1925 \cite{debroglie1925,debroglie1925b} that such a time-symmetric field could be central to justify wave-particle duality and the stability of the non-radiating Bohr's orbits in atoms.  Not surprisingly, time-symmetry can be used to avoid some paradoxical consequences of Bell's theorem \cite{Bell} concerning the non-locality  of quantum mechanics. Here, we will show indeed that our time-symmetric version of  de Broglie DS can interpret the nonlocality of quantum mechanics as resulting from a much deeper non-linear but local dynamics involving a time-symmetric field (i.e., involving a half sum of retarded and advanced fields).   In turn,  we show that this approach in the four dimensional space-time exactly recovers the dBB interpretation of de Broglie and Bohm for the entangled motions of quantum particles in the configuration space. Therefore, the  DS model we propose here is able to recover quantum mechanics for spinless particles coupled to external fields. Altogether this work also bring new insights in the context of quantum hydrodynamic analogs \cite{Couder,Bush} or related mechanical proposals made by the author and collaborators~\cite{Jamet1,Jamet2}.\\
\indent The plot of the article is the following: In Section \ref{sec2} we review the content of Bell's theorem and the challenge it puts on the notion of locality , causality and statistical indepedence. In section \ref{sec3} we give a precise description of the the time-symmetric DS theory involving a non-linear wave equation for soliton (this work extend a previous related analysis \cite{Drezet2023}). In particular we stress the role of the `phase harmony' condition and explain how to recover Bohmian mechanics and the famous `guidance formula' out of this theory. In Section \ref{sec4} we discuss the many-body problem involving $N$ entangled solitons. We show how this $N$ solitons or singularities are driven by a Schr\"odinger of Klein-Gordon linear pilot wave-field $\Psi$ allowing us to recover the consequences of the dBB pilot wave theory. Finally, in the discussion/conclusion section \ref{sec5} we discuss how our model evades the usual nonlocality of the standard dBB pilot-wave theory and replaces it by a superdeterministic link driven by the time-symmetry of our fundamental de Broglie nonlinear field.    
\section{Motivations: The nonlocality conundrum}\label{sec2}
\indent Bell's theorem \cite{Bell} is probably one of the most important results in the Physics of the XX$^{th}$ century. Briefly stated, it shows that any explanations of quantum mechanics involving or not hidden variables must be `non-local'. Actually, this is an oversimplification since the theorem  involves several fundamental axioms concerning locality, causality, and statistical independence that must be fulfilled in order to derive the famous Bell's inequalities.\\ \indent Let us review that issue briefly. Starting with a pair of entangled particles 1 and 2 prepared in a source $S$ and subsequently (space-like) separated  in two remote regions $A$  and $B$, Bell considers the joint quantum probability $P_{12}(\alpha,\beta|\mathbf{\underline{a}},\mathbf{\underline{b}})$  for observing  particle 1 with the property $\alpha=\pm 1$ and particle 2 with the property $\beta=\pm 1$ (e.g.,  associated with spin observables) granted that the settings of the measuring apparatus in region A and B are prepared as $\mathbf{\underline{a}}$ and $\mathbf{\underline{b}}$ respectively. Assuming the existence of hidden variables or more generally beables  $\boldsymbol{\Lambda}$ in order to describe quantum mechanics we can write 
\begin{eqnarray}
P_{12}(\alpha,\beta|\mathbf{\underline{a}},\mathbf{\underline{b}})=\int dP_{12}(\alpha,\beta,\boldsymbol{\Lambda}|\mathbf{\underline{a}},\mathbf{\underline{b}})=\int P_{12}(\alpha|\beta,\boldsymbol{\Lambda},\mathbf{\underline{a}},\mathbf{\underline{b}})P_{12}(\beta|\boldsymbol{\Lambda},\mathbf{\underline{a}},\mathbf{\underline{b}})dP_{12}(\boldsymbol{\Lambda}|\mathbf{\underline{a}},\mathbf{\underline{b}})
\end{eqnarray}    where the first equality means that the experimental probability $P_{12}(\alpha,\beta|\mathbf{\underline{a}},\mathbf{\underline{b}})$ is recovered by summing over the different actualisations of the variables (Bell beables) $\boldsymbol{\Lambda}$ associated with the particles, and the second equality  is just an application of the mathematical definition of conditional probablities.  In his derivation  Bell further assumed that:   
\begin{eqnarray}
P_{12}(\alpha|\beta,\boldsymbol{\Lambda},\mathbf{\underline{a}},\mathbf{\underline{b}})=P_{12}(\alpha|\boldsymbol{\Lambda},\mathbf{\underline{a}})\label{a}\\
P_{12}(\beta|\boldsymbol{\Lambda},\mathbf{\underline{a}},\mathbf{\underline{b}})=P_{12}(\beta|\boldsymbol{\Lambda},\mathbf{\underline{b}})\label{b}\\
dP_{12}(\boldsymbol{\Lambda}|\mathbf{\underline{a}},\mathbf{\underline{b}})=dP_{12}(\boldsymbol{\Lambda})\label{c}
\end{eqnarray}
The two Eqs.~2 and 3 are not controversial and are associated with the notion of local-causality already advocated by Einstein. The idea is that by assuming the observers Alice and Bob in regions A and B are choosing `freely'  and at the last moments  the  setting directions $\mathbf{\underline{a}}$ and $\mathbf{\underline{b}}$ it must (according to the principle of Einstein's special relativity) be impossible to have faster than light (tachyonic) propagation of any physical influence from A to B or B to A affecting the `independent' measurements. The measurements done by Alice and Bob must therefore only depend on the local properties $\mathbf{\underline{a}}$ or $\mathbf{\underline{b}}$ and on the shared hiddden variable $\boldsymbol{\Lambda}$ associated with the `preparation at the source' (common past) of the two entangled particles. The last Eq.~4 is even more obvious and natural and states that the hidden variables  prepared at the source  $S$ must be independent of the settings $\mathbf{\underline{a}}$ and $\mathbf{\underline{b}}$ because the choice was made freely at the last moment after the preparation of the entangled pair at the source. Relaxing this statistical indepedence assumption would apriori leads to a superdeterministic and conspiratorial Universe that seems to conflict with the goal and methodology of science.  \\
\indent  With all these natural hypotheses  Bell proved that a specifical statistical property written $\langle S(\mathbf{\underline{a}},\mathbf{\underline{b}},\mathbf{\underline{a}}',\mathbf{\underline{b}}')\rangle$ and depending on different possible choices of the settings $\mathbf{\underline{a}},\mathbf{\underline{a}}'$ at $A$ and $\mathbf{\underline{b}},\mathbf{\underline{b}}'$ at $B$ must be bound: $\langle S(\mathbf{\underline{a}},\mathbf{\underline{b}},\mathbf{\underline{a}}',\mathbf{\underline{b}}')\rangle\leq 2$. The problem is that quantum mechanics predicts and experiments show in some cases Bell's inequalities is violated  up to the value $\langle S_{\textrm{Tsirelson}}(\mathbf{\underline{a}},\mathbf{\underline{b}},\mathbf{\underline{a}}',\mathbf{\underline{b}}')\rangle=2\sqrt{2}$. In other words the theorem reveals a  logical contradiction arizing from the simultaneaous supposed validity of quantum mechanics and the existence of locally-causal statistically independent hidden variables. Something must be wrong and assuming  quantum mechanics is true this implies that at least one of Bell's hypotheses must be abandonned. Furthermore, assuming the existence of hidden variables (i.e., assuming a strong form of realism  that refuses a bare operationalist/positivist approach) implies necessarily that we  must relax at least one of the conditions 2-4. In this context it is useful to remind that the historical experiments of Aspect et al. \cite{Aspect} closed the `locality loophole' using periodical switching devices, and subsequent experiments \cite{Zeilinger} used `genuine' quantum-random-number generators (i.e., single photons sources) assuming no-superderminism. The `detection loophole' can be closed in some cases \cite{Hanson}, and  some tests also excluded `tachyonic loopholes' (with some assumptions) leading to the lower limit for the velocity of non-local information of $\sim 50000$ times the velocity of light \cite{Gisin} (the tests assume that the signal is propagating in the future and cannot refute the dBB theory that involves necessarily signals travelling forward or backward in time in different reference frames\cite{DrezetFP2019}). Moreover, recently closing the `freedom of choice' loophole was realized using switching devices monitored by photons emitted $\sim$10 Billions years ago by quasars \cite{Zeilinger2,Zeilinger3,Kaiser,Zeilinger4}. Assuming that there is no nonlocality (i.e., no instantaneous or tachyonic action at a distance, including dBB connections) this lets the conspiratorial or super/hyper deterministic loophole as the only serious reminding loophole:  Indeed, such `Cosmic Bell' correlations seem to imply everything should be fine-tuned and conspiratorially correlated from more or less the Big-Bang time in order to reproduce quantum predictions \cite{Hooft,Kaiser} [Note moreover that cosmic inflation is supposed to save causality without superdeterminism by providing an explanation for the homogeneity of the  cosmological microwave background].  \\
\indent Moreover, Bell's understood very well from the start \cite{Bell} that de Broglie in 1927 \cite{debroglie1927,Valentini} and Bohm in 1952 \cite{Bohm1952,Hiley} already developed a rigorous deterministic, and explicitly non-local hidden variable theory. In this dBB approach the particles are  point-like objects guided by the entangled wavefunction creating a nonlocal link between the particles. That is, in the dBB pilot-wave theory the two apparently natural relations Eqs.~2 and 3 don't hold true ($\boldsymbol{\Lambda}$ is now associated with the spatial coordinates of the particles in the remote past at the emission time by the source $S$). This means that some kind of instantaneous action-at-a-distance  exists between the particles and therefore the measurements are not really independent (even if this cannot be used to send `macroscopic' faster than light signals). Bell's following Bohm therefore acknowledged this remarkable and elegant dBB theory that is curiously in tension with the spirit of special relativity but that  nevertheless `peacefully' hiddes the tachyonic effects at the microscopic level of the hidden variables in such a way as to reproduce exactly the statistical prediction of quantum mechanics. Importantly, the dBB theory assumes statistical independence and as a consequence Eq.~4 still holds true. Moreover, not everyone is pleased with the nonlocality of dBB theory entailing necessarily a preferred reference frame or space-time foliation looking like a reminiscence of the prerelativistic era and its `Aether' substratum \cite{Hiley}. Yet, the dBB theory is devoid of any logical contradiction involving tachyonic signals (i.e., like influencing its own past to create a forbidden causal loop or paradox \cite{DrezetFP2019}), and the theory can also be generalized by associating an hidden variable to the preferred foliation $\mathcal{F}$ of space-time  specifying the particle dynamics (i.e., in order to recover a democracy and symmetry between the different folliations without introducing an Aether \cite{DrezetFP2019}). Nevertheless, the dBB theory still looks odd and counterintuitive. For these reasons and others many authors attempted different approaches like  Everett's Many-Worlds interpretation \cite{Everett}. Indeed, it is sometimes claimed that Everett's theory is not in tension with special relativity because the theory avoids the `single-world' picture associated with hidden variables theories; however, it can be shown that such an unfounded statement is based on overlooking the status of probability that can not be defined unambiguously in the Many-Worlds theory~\cite{Drezetsymmetry}. Therefore, in the following we will not consider such attempts.
\section{The time-symmetric double solution program}\label{sec3}
\subsection{The soliton near-field}\label{sub31}
\indent In this work we  consider a particular development of the dBB theory: Namely the DS theory proposed originally by de Broglie in 1925 \cite{debroglie1925,debroglie1925b}. This theory, we will show, can bring new insights concerning the issue of non-locality. Moreover, it is important to mention that  de Broglie strongly modified his theory in 1927 and in the 1950's \cite{debroglie1927,debroglie1956} in collaboration with Vigier \cite{Vigierthesis}. It is often this last version that is mentionned in the literature, i.e., when it is not completely ignored (for useful reviews about the DS approach see \cite{Fargue,DrezetReview}). The DS theory, inspired from Einstein early works on photons, is based on the idea to describe particles as moving singularities of a classical scalar field theory (by a singularity we mean that the field is infinite at the position of the particle). De Broglie first considered \cite{debroglie1925,debroglie1925b}  that each particle is  a point-like moving clock pulsating at its Compton frequency $\omega_0=mc^2/\hbar$ (in the rest of this work we will use the relativistic conventions  $c,\hbar=1$). This actually means that the particle has internal properties.  If this particle is at rest in he laboratory frame the clock `generates' an extended stationnary field surrounding the singularity. When the particle is in motion this field guides the particles and interacts with obstacles.  De Broglie hoped  \cite{debroglie1925b} that the interaction between this wave-like field, the particle and its environement could explain wave-particle duality (associated with interference fringes) and the stationnary orbits required in the Bohr atomic model .\\
\indent The fundamental scalar field  $u(t,\mathbf{x}):=u(x)\in \mathbb{C}$ used by de Broglie is outside the singularities at space-time point $x:=[t,\mathbf{x}]\in\mathbb{R}^4$ a solution of a basic linear and local field equation, i.e.:
\begin{eqnarray}
\Box u(x)=\partial^2 u(x)=(\partial_t^2-\boldsymbol{\nabla}^2)u(x)=0. \label{Dal}
\end{eqnarray}
The simplest oscillating solution associated with the pulsation $\omega_0$ defined in the rest frame $\mathcal{R}$ of the singularity is the monopolar (spherically symmetric) field:
\begin{eqnarray}
u(x)=\frac{g_0}{4\pi}e^{-i\omega_0 t}\frac{\cos{(\omega_0 R)}}{R}\label{oldone}
\end{eqnarray} with $R=|\mathbf{x}|$ the radial distance to the singularity (particle) located at the spacial origin, and $g_0$ a constant. Remarkably, when this monopole field is studied from a different Lorentz frame $\mathcal{R}'$ where the field singularity moves at the velocity $v$ (along the $x$-direction) the relativistically invariant scalar $u-$fields reads now
\begin{eqnarray}
u(x)=u'(x')=\frac{g_0}{4\pi}e^{-i\omega_0 \gamma(t'-vx')}\frac{\cos{(\omega_0 R)}}{R}
\end{eqnarray} with $R=\sqrt{(y'^2+z'^2+\gamma^2(x'-vt')^2}$ and $\gamma=1/\sqrt{(1-v^2)}$. What is beautiful here is that the (scalar relativistic invariant) phase wave $\Psi=e^{-i\omega_0 \gamma(t'-vx')}$ involved in $u$ is actually a plane wave solution of the linear Klein-Gordon equation $\Box \Psi(x)=-\omega_0^2 \Psi(x)$. It is easily shown that the particle (clock) is synchronized with the $\Psi-$wave and therefore the $u-$field during its motion. The whole picture is interesting because in the one hand it is inherently classical and relativistic, i.e., in the same sense as Maxwell theory or general relativity are classical and relativistic (the $u-$field propagates in space-time with the particle). On the other hand, this approach suggests that wave-particle duality and more generally quantum mechanics is just a sophisticated version of classical physics involving oscillating and moving point sources coupled to a field. In order to prove this hypothesis de Broglie hoped to be able to calculate trajectories of the singularities in complex environments involving external potentials like the Coulomb potential or the double slit barrier. In the DS approach the $u-$field should be able to influence the motion of the particle singularities in such a way as to reproduce quantum mechanics. We stress that in 1927 de Broglie used the linear Klein-Gordon equation instead of the simple Eq.~\ref{Dal} but this doesn't simplify the analysis \cite{debroglie1927,DrezetReview}.\\
\indent  Clearly, this picture with trajectories guided by a wave is linked to the dBB pilot-wave theory discussed before. In the 1920's de Broglie could not develop this DS program mathematically (see the discussion in \cite{DrezetReview}) and it is mostly for this reason that he switched to the simpler dBB pilot-wave approach.  The problem of course is that in the dBB theory the nature of the guiding field is more obscure and more `epistemic'.  In the end, the dBB theory was developped in the configuration space not in the 3D physical space and therefore the initial intuition of the DS project was lost. Importantly de Broglie also abandonned in 1928 the pilot-wave theory because he felt the theory was unable to explain the strong form of nonlocality involved in quantum mechanics (in 1928  that issue was related to the difficulty for understanding the mysterious concept of `wave function collapse' or `reduction' in term of Einstein special relativity prohibiting faster than light communication~\cite{debroglie1930}).  Moreover, in 1952 de Broglie together with Vigier \cite{debroglie1956,Vigierthesis} and few collaborators came back to the DS program and tried to show that singularities or solitonic solutions of some unknown wave equations are following the paths or trajectories predicted by the dBB theory in the configuration space. The most important modification they suggested  was, in analogy with general relativity, to consider nonlinear wave equations for the $u-$field in order to remove the mathematical singularities at the particle positions. In such an approach particles are becoming `solitons', i.e., localized solitary `bunched' waves propagating as a whole without dispersion.  However, nonlinear wave equations are even more difficult to solve than linear wave equations with moving singularities. Beside that point they could not define a  precise nonlinear wave equation that could implement the DS goals.  Moreover,  the most problematic point is perhaps that de Broglie underestimated the  impact of nonlocality in the DS theory.  Indeed, the dBB  approach is non-local in the configuration space.   How a nonlinear but local classical field theory defined in the 4D relativistic space-time could reproduce and justify the non-locality of the dBB theory?  Following some old intuitions going back to Einstein they suggested that nonlocality was existing only in the interacting regime when the particle are not too much separated \cite{dasilva,debroglie1971}. Furthermore, Bell's theorem came in 1965 and de Broglie couldn't assimilate and accept the lesson of this central result \cite{debroglie1974,Bellreply} (that was not the case of Vigier who stopped working on the DS project for a while and moved to alternative nonlocal and stochastic hidden variables approaches \`a la Bohm or Nelson~\cite{Vigier}). \\
\indent   In the present research (discussed in more details in \cite{Drezet2023} following a related article \cite{Drezet2023b}) we are tooking seriously the idea of a particle-soliton but at the same time we are going back to the old 1925 ideas sketched in Eqs.~5-7. What we found motivating is the deep time-symmetry of this old picture.   Indeed, it is visible that the field  Eq.~6 is a solution of the wave equation $\Box u(x)=g_0\delta^{3}(\mathbf{x})e^{-i\omega_0 t}$ involving a source term.  Eq.~\ref{oldone} is actually the time-symmetric solution, i.e., the half-sum of a retarded and advanced radiating waves: \begin{eqnarray}
u_{\textrm{ret}}(x)=\frac{g_0}{4\pi}e^{-i\omega_0 t}\frac{e^{i\omega_0 R}}{R}\nonumber\\
u_{\textrm{adv}}(x)=\frac{g_0}{4\pi}e^{-i\omega_0 t}\frac{e^{-i\omega_0 R}}{R}\nonumber\\
u(x)=\frac{1}{2}[u_{\textrm{ret}}(x)+u_{\textrm{adv}}(x))].
\end{eqnarray}
What actually inspired de Broglie  was the work of Tetrode and Page~\cite{Tetrode,Page}  (later rediscovered by Fokker~\cite{Fokker}, and Wheeler with Feynman~\cite{Wheeler}) on the time-symmetric electrodynamics in which action at a distance is mediated by the half-sum of retarded and advanced electromagnetic waves emitted by electrons and protons.  Page~\cite{Page} suggested that the time-symmetric field could explain why electrons orbits in Bohr's atoms do not radiate and this was one of the initial crux of de Broglie attempt in 1925 \cite{debroglie1925,debroglie1925b}. Interestingly, de Broglie  abandonned this idea in 1926  and never came back to it even after his former student Costa de Beauregard rediscovered a sequel of the idea in 1942 \cite{Costa} in order to solve the EPR paradox! After this work \cite{Costa} many retrocausal and time-symmetric theories were developed in order to explain the violations of Bell's inequalities (see for example \cite{Costa2,Cramer,Aharonov} and the interesting retrocausal  dBB theory \cite{Sutherland,Sen}). All this clearly motivates the present work. \\ 
\indent Here, we start from  de Broglie DS and postulate the following nonlinear wave equation for the $u-$field \cite{Drezet2023}:
\begin{eqnarray}
D^2u(x)=\frac{3l_0^2}{(\frac{g_0}{4\pi})^4}(u(x)u^\ast(x))^2u(x)\label{1b}\nonumber\\
\end{eqnarray} where $D=\partial+ieA(x)$ is a covariant derivative  involving the electromagnetic potential four-vector $A(x):=[V(x),\mathbf{A}(x)]$ and $e$ an electric charge. The nonlinear function we consider \cite{Drezet2023} is the simple `Lane-Emden' fifth power law where  $g_0$ is a (nondimensional) coupling constant and $l_0$ a length  that will define the typical radius  of our soliton. Importantly, this nonlinearity allows us to define explicit analytical solitons (at least in the near-field of the particle, i.e., for distances $R$ to the center such that $R,l_0\ll \omega_0^{-1}$). This equations is clearly different from the quantum linear Klein-Gordon equation $D^2\Psi(x)=-\omega_0^2\Psi(x)$ involving the standard relativistic wavefunction $\Psi(x)$ for scalar (spin 0) particle. We stress  that if the field decays sufficiently we can in the far-field of the soliton   approximates Eq.~\ref{1b}  by the linear equation $D^2u(x)\simeq 0$. This property is fundamental since it allows us to develop a simplified approach for describing the soliton if $R\gg l_0$ (as shown in Section \ref{sub32}).   \\ 
\indent In order to find a solution of Eq.~\ref{1b} we use the polar form $u(x)=f(x)e^{i\varphi(x)}$ and obtain after separation:
\begin{subequations}
\label{NL}
\begin{eqnarray}
(\partial \varphi(x)+eA(x))^2=-\frac{3l_0^2}{(\frac{g_0}{4\pi})^4}f^4(x))+\frac{\Box f(x)}{f(x)}:=\mathcal{M}^2_u(x)\label{2c}\\
\partial[f^2(x)(\partial \varphi(x)+eA(x))]=0.\label{2d}
\end{eqnarray}
\end{subequations} Eq.~\ref{2c} is generally named the (nonlinear) dBB Hamilton-Jacobi equation, and Eq.~\ref{2d} is reminiscent of the electric current conservation. This defines an hydrodynamical representation of the non-linear Klein-Gordon equation. Similarly, we can define an hydrodynamical representation for the field $\Psi(x)=a(x)e^{iS(x)}$ solution of the linear Klein-Gordon equation: 
\begin{subequations}
\begin{eqnarray}
(\partial S(x)+eA(x))^2=\omega_0^2+Q_\Psi(x):=\mathcal{M}^2_\Psi(x) \label{2}\\
\partial[a^2(x)(\partial S(x)+eA(x))]=0,\label{2b}
\end{eqnarray}
\end{subequations} with $Q_\Psi(x)=\frac{\Box a(x)}{a(x)}$ the so called quantum potential \cite{debroglie1927,Bohm1952,debroglie1956} used in the pilot-wave theory.
 We solve the pair of equations \ref{NL} in the region where the soliton profile is supposed to be well peaked (i.e., near the center or soliton `core' surrounding a mean trajectory $z(\tau)$ labeled by the proper time $\tau$ along the path). For this we assume with de Broglie the so-called `phase-harmony condition'~\cite{debroglie1927,debroglie1956}:
\begin{quote}
\textit{To every regular solution $\Psi(x)=a(x)e^{iS(x)}$ of the linear Klein-Gordon equation corresponds a localized solution $u(x)=f(x)e^{i\varphi(x)}$ of Eq.~\ref{1b} having \underline{locally} the same phase $\varphi(x)\simeq S(x)$, but with an amplitude $f(x)$ involving a generally moving soliton centered on the path $z(\tau)$ and which is representing the particle.}\end{quote} In other words, this phase-locking condition forces the two waves $u$ and $\Psi$ to vibrate locally in unison. We thus consider a Taylor expansion of the phase $\varphi(x)$ in the vicinity of $z(\tau)$ in a space like hyperplane crossing $z(\tau)$ and defining a local instantaneous rest frame for the particle center (this hyperplane $\Sigma(\tau)$ is defined by the condition $\xi_\mu\dot{z}^\mu(\tau):=\xi\cdot\dot{z}(\tau)=0$ with $\xi=x-z(\tau)$). We assume: 
\begin{eqnarray}
\varphi(x) \simeq S(z(\tau)) -eA(z(\tau))\xi+B(z(\tau))\frac{\xi^2}{2}+O(\xi^3)\label{phaseharmony}
\end{eqnarray} where $B(z(\tau))$ is a new collective coordinate introduced to increase the number of degrees of freedom. As shown   in \cite{Drezet2023,Drezet2023b} we can define the fluid velocities  $v_u(x)=-\frac{\partial \varphi(x)+eA(x)}{\mathcal{M}_u(x)}$ and $v_\Psi(x)=-\frac{\partial S(x)+eA(x)}{\mathcal{M}_\Psi(x)}$ in the regions where $\mathcal{M}^2_u(x),\mathcal{M}^2_\Psi(x)\geq 0$. The velocity $v_\Psi(x)$ is associated with the dBB pilot-wave interpretation of 1927~\cite{debroglie1927} and de Broglie postulated that the particle his guided by the $\Psi-$wave: 
\begin{eqnarray}
\frac{d z(\tau)}{d\tau}=v_\Psi(z(\tau))=-\frac{\partial S(z(\tau))+eA(z(\tau))}{\mathcal{M}_\Psi(z(\tau))}\label{PWIguidance}
\end{eqnarray} This formula is non ambiguous at least in the regime $\mathcal{M}^2_\Psi(x)>0$ avoiding tachyonic trajectories and we limit our analysis to that case in the main text of this article. However, the tachyonic or superluminal regime can be self-consistently  described in the dBB pilot-wave approach and is discussed in the context of our DS proposal in the Appendix \ref{appendix}. Moreover, from our DS theory and Eq.~\ref{phaseharmony} we deduce \cite{Drezet2023,Drezet2023b} $v_u(x)\simeq \frac{d z(\tau)}{d\tau} + O(\xi)$ and similarly $\mathcal{M}_u(x)\simeq \mathcal{M}_\Psi(z(\tau))+\boldsymbol{\xi}\cdot\boldsymbol{\nabla}\mathcal{M}_\Psi(z(\tau))+O(\xi^2)$, $\partial_\mu\mathcal{M}_u(z(\tau))=\partial_\mu\mathcal{M}_\Psi(z(\tau))$. The whole picture is actually self consistent if the soliton central trajectory $z(\tau)$ is identified with the dBB pilot-wave trajectory given by Eq.~\ref{PWIguidance}.  We have thus 
\begin{eqnarray}
\frac{d z(\tau)}{d\tau}=v_\Psi(z(\tau))=v_u(z(\tau))\label{DSguidance}
\end{eqnarray} justifying  the guidance postulate of de Broglie. In other words, we show that it is always possible to find locally a first-order matching $\varphi(z(\tau))=S(z(\tau))$, $\partial_x \varphi(x)|_{x=z(\tau)}=\partial_z S(z(\tau))$. The two phase waves $\varphi$ and $S$ of the two fields $u$ and $\Psi$ are thus connected along the curve $z(\tau)$ and this is the core of the DS or phase harmony approach. We stress that from  the dBB  dynamics Eqs. \ref{2}, \ref{2b} and the guidance formula we get the second-order relativistic `Newton' law already found by de Broglie in 1927\cite{debroglie1927CR} \begin{eqnarray}
\frac{d}{d\tau}[\mathcal{M}_\Psi(z(\tau))\dot{z}^\mu(\tau)]=\partial^\mu [\mathcal{M}_\Psi(z(\tau))]
+eF^{\mu\nu}(z(\tau))\dot{z}_{\nu}(\tau)
 \label{Newton}
\end{eqnarray}  with $F^{\mu\nu}(x)=\partial^\mu A^\nu(x)-\partial^\nu A^\mu(x)$ the Maxwell tensor field at point $x:=z$. The varying de Broglie mass $\mathcal{M}_\Psi(z(\tau))$ (i.e., varying quantum potential $Q_\Psi(z(\tau))$) is central in order to recover the non-classical features of quantum mechanics specific of the dBB pilot-wave theory.\\
\indent An important relation is obtained from the hydrodynamical  conservation  Eq.~\ref{2d} and the phase-harmony Eq.~\ref{phaseharmony} constraint:
\begin{eqnarray}
-\partial v_u(z(\tau)):=-\frac{d}{d\tau}\ln{[\delta^3\sigma_0(z(\tau))]}
=\frac{d}{d\tau}\ln{[f^2(z(\tau))\mathcal{M}_\Psi(z(\tau))]}=\frac{3B(z(\tau))}{\mathcal{M}_\Psi(z(\tau))}\label{newdfluid}
\end{eqnarray} The first two equalities concerns the fluid deformation and compressibility where we introduced an infinitesimal comoving 3D fluid volume $\delta^3\sigma_0$ (defined in the local rest frame $\mathcal{R}_\tau$ associated with $\Sigma(\tau)$) driven by the fluid motion. In particular, we see that if the soliton is undeformable we must have $-\partial v_u(z(\tau)):=-\frac{d}{d\tau}\ln{[\delta^3\sigma_0(z(\tau))]}=0$ and thus from the last equality in Eq.~\ref{newdfluid} we get  $B(\tau)=0$. Moreover, if this soliton is undeformable we must also have $f(z(\tau))=Const.$ $\forall \tau$ and thus from Eq.~\ref{newdfluid} $\mathcal{M}_\Psi(z(\tau))=Const.$ must hold along the trajectory $z(\tau)$. This is conflicting with the dBB theory imposing a varying $Q_\Psi(z(\tau))$ along  particle trajectories. We conclude that we will have to relax the natural assumption of undeformability (i.e., $\delta^3\sigma_0(z(\tau))=Const.$, $f(z(\tau))=f_0=Const.$) in order to develop a self-consistent DS theory reproducing the dBB trajectories, i.e., agreeing with quantum predictions.\\
\indent In order to find the soliton profile $f$  we rewrite Eq.~\ref{2d} in the rest-frame $\mathcal{R}_\tau$ near the soliton center (in the Fermi limit $\xi\ddot{z}\ll 1$ where  we have $|\partial_t^2f|\ll|\boldsymbol{\nabla}^2f|$ \cite{Drezet2023,Drezet2023b})  and obtain  the partial differential equation for the soliton profile for points $x$ belonging to $\Sigma(\tau)$ and localized near $z(\tau)$:
\begin{eqnarray}
\mathcal{M}^2_\Psi(z(\tau))f(x)+\boldsymbol{\nabla}^2f(x)\simeq -\frac{3l_0^2}{(\frac{g_0}{4\pi})^4}f^5(x))
\label{ODE}\end{eqnarray} with $\boldsymbol{\nabla}:=\frac{\partial}{\partial\boldsymbol{\xi}}$ \cite{Drezet2023,Drezet2023b}.  Furthermore, in the near-field if we suppose the soliton core size $l_0$ to be much smaller than the Compton wavelength  $\omega_0^{-1}\sim \mathcal{M}_\Psi^{-1}$ we can use the stronger approximation
 \begin{eqnarray}
\boldsymbol{\nabla}^2f(x)\simeq -\frac{3l_0^2}{(\frac{g_0}{4\pi})^4}f^5(x)
\label{ODEnf}\end{eqnarray} which is known as the Lane-Emden equation\cite{Drezet2023}. This equation admits the spherically symmetrical exact-solution:
\begin{eqnarray}
f(x):=F_\alpha(r)=\frac{\sqrt{\alpha}g_0}{4\pi}\frac{1}{\sqrt{\alpha^2r^2+l_0^2}}=\frac{g_0}{4\pi \sqrt{\alpha}}\frac{1}{\sqrt{r^2+\frac{l_0^2}{\alpha^2}}}\label{Lanesolutionscaled}
\end{eqnarray} which is parametrized by the constant $\alpha\in \mathbb{R}^+$ and has the dilation invariance $F_\alpha(r)=\sqrt{\alpha}F_1(\alpha r)$. Far away from the soliton `core', i.e., if $r\gg l_0$, this field has the asymptotic monopolar limit $F_\alpha(r)\simeq \frac{g_0}{4\pi \sqrt{\alpha}}\frac{1}{r}$. In this limit it is justified to use instead Poisson's equation $\boldsymbol{\nabla}^2f=-\frac{g_0}{\sqrt{\alpha}}\delta^3(\textbf{x})$ for a point-like source with effective `scalar-charge' $\frac{g_0}{\sqrt{\alpha}}$.  We will come back to the assymptotic field later but for the moment we stress that Eq.~\ref{Lanesolutionscaled} is an approximate equation keeping its general validity if the motion of the soliton is not varying too fast (as justifed in \cite{Drezet2023b}) and  we thus  physically interpret the parameter $\alpha$ as a new collective coordinate for the soliton.  More precisely, we now assume that during its motion the soliton typical extension $l(z(\tau))$ changes(adiabatically) with time $\tau$ and we write
\begin{eqnarray}
l(z(\tau))=l_0/\alpha(\tau),&
g(z(\tau))=\frac{g_0}{\sqrt{\alpha(\tau)}}
\end{eqnarray} or equivalently
\begin{eqnarray}
l(z(\tau))=l(z(0))\alpha(0)/\alpha(\tau),&
g(z(\tau))=\frac{g(0)}{\sqrt{\alpha(\tau)/\alpha(0)}}\label{const}
\end{eqnarray}
 where  $\alpha(\tau)$ defines the dynamics concerning the radius (the proper time $\tau=0$ is chosen arbitrarily to correspond to an initial point $z(0)$ along the trajectory).\\
\indent In order to fix the $\alpha$-dynamics we use the 
local conservation law for a fluid element located at the soliton center Eq.~\ref{newdfluid}: $\frac{d}{d\tau}\log{[f^2(z(\tau))\mathcal{M}_\Psi(z(\tau))\delta^3\sigma_0(z(\tau))]}=0$  and  we have   by integration 
$f^2(z(\tau))\mathcal{M}_\Psi(z(\tau))\delta^3\sigma_0(z(\tau))=f^2(z(0))\mathcal{M}_\Psi(z(0))\delta^3\sigma_0(z(0))$ defining a constant of motion along a given trajectory. 
Furthermore, from Eq.~\ref{Lanesolutionscaled} $f(z(\tau))=F_{\alpha(\tau)}(0)=\sqrt{\alpha(\tau)}F_1(0)=\sqrt{\alpha(\tau)/\alpha(0)}f(z(0)$ and $\delta^3\sigma_0(\tau)=\frac{\alpha^3(0)}{\alpha^3(\tau)}\delta^3\sigma_0(\tau)$ (a more rigorous justification is given in \cite{Drezet2023}).  and  we deduce:
\begin{eqnarray}
\alpha(\tau)=\alpha(0)\sqrt{\frac{\mathcal{M}_\Psi(z(\tau))}{\mathcal{M}_\Psi(z(0))}}. \label{inteF}
\end{eqnarray}  
Moreover, from Eqs.\ref{newdfluid} and \ref{inteF} we deduce \begin{eqnarray}
B(z(\tau))=\frac{1}{2}\frac{d}{d\tau}\mathcal{M}_\Psi(z(\tau)) \label{inteG}
\end{eqnarray}  which together with Eq. \ref{inteF} defines the complete deformation/compression of the soliton near-field.\\
\indent We stress that the model developed here for a subluminal soliton can be extended to the superluminal or tachyonic sector. This is presented briefly in the Appendix \ref{appendix}. While it could at first look curious to derive superluminal motions from a purely local DS theory respecting the principle of special relativity we cannot apriori forbid such a regime since the dBB pilot-wave theory for a point-like  particle obeying the Klein-Gordon theory predicts that in some cases such a particle can reach the speed of light and even cross the light cone (i.e., if and only if  the mass $\mathcal{M}_\Psi(z)$ vanishes while the particle crosses the light cone). In the context of the DS theory, trying to reproduce the dBB pilot-wave predictions,  the existence of tachyonic waves is reminiscent of the so called `X-waves' observed  in optics where a region of a  wave-packet is allowed to propagate faster than the speed of light in vacuum if and only if this cannot be used to transfer information or energy, i.e., similar to a phase-wave (see \cite{Saari} for an illuminating discussion on this non-signalling constraint on X-waves).  Since the tachyonic dBB motions are generally associated with evanescent or transient $\Psi-$fields localized in finite space-time regions there is no way to violate no-signalling.  Using the DS theory the solutions we obtain also involve some `X-waves' which can only have a physical meaning in transient regions. Therefore, altogether the picture obtained from the DS theory is consistent even in the tachyonic regime.       
\subsection{The soliton far-field}
\label{sub32}
\indent The previous theory developed for the near-field can be used to define the mide-field and far-field of the soliton i.e., if we dont neglect the mass term $\mathcal{M}_\psi(z(\tau))$ in Eq.~\ref{ODE}. We consider first the case of an uniform motion  where $\mathcal{M}_\Psi=Const.=\omega$ and search for a spherical solution of 
\begin{eqnarray}
\frac{d^2}{dr^2}F(r)+\frac{2}{r}\frac{d}{dr}F(r)+\frac{3l_0^2}{(\frac{g_0}{4\pi})^4}F^5(r)+\omega^2 F(r)=0.
\label{Laneradial}
\end{eqnarray}
As shown in \cite{Drezet2023} if we can assume $\omega l_0\ll 1$ (i.e., a very small soliton) Eq.~\ref{Laneradial} admits the solution\begin{eqnarray}
F_\alpha(r)\simeq\frac{\sqrt{\alpha}g_0}{4\pi}\frac{\cos{(\omega r)}}{\sqrt{\alpha^2r^2+l_0^2}}\label{interpolate1}
\end{eqnarray}  which is an interpolation between the near-field Eq.~\ref{Lanesolutionscaled} and the far-field de Broglie monopolar solution 
 \begin{eqnarray}
F_\alpha(r)\simeq\frac{g_0}{4\pi\sqrt{\alpha}}\frac{\cos{(\omega r)}}{r}\label{ODEFF}
\end{eqnarray} obtained if $\omega r\gg 1$. Such a far-field is a solution of the inhomogeneous d'Alembert equation $\Box u(t,\textbf{x})=\frac{g_0}{\sqrt{\alpha}}\delta^{3}(\textbf{x})e^{-i\omega t}$ with $u(t,\textbf{x})=e^{-i\omega t}F_\alpha(r)$.\\
\indent Two remarks can be done here:  First, note that the coefficient $\alpha$ and the frequency $\omega$ of the soliton are not determined univocally by the theory.  The soliton admits a continuous  frequency spectrum corresponding to a continuous mass spectrum $\omega:=\omega_0\in [0,+\infty[$. Hence, it means that the present theory would have to be completed  to fix the mass of the particle.  Moreover, assume  that the mass $\omega_0$ is fixed. In the context of the dBB pilot-wave theory (which in our DS approach defines the soliton trajectory) it corresponds to a situation where from Eq.\ref{2} $Q_\Psi=0$ (i.e., $\mathcal{M}_\Psi=\omega=\sqrt{(\omega_0^2+Q_\Psi)}=\omega_0$). This is  associated with a guiding plane wave $\Psi(x)=A e^{-ikx}$ which in the rest frame reads $\Psi(x)=A e^{-i\omega_0 t}$.   Yet, the dBB pilot-wave theory can also generate constant masses  different from $\omega_0$  if the quantum potential $Q_\Psi$ is constant but different from zero. This is for example the case  if the guiding field solution of the linear Klein-Gordon equation reads $\Psi(t,x)=A\cos{(kx)}e^{-i\omega t}$ (corresponding to a 1 dimensional stationnary wave along the spatial $x$ direction with pulsation $\omega=\sqrt{(\omega_0^2+k^2)}> \omega_0$ and $k$ the wavevector along the $x$ direction) or  $\Psi(t,x)=Ae^{\pm\kappa x }e^{-i\omega t}$ (corresponding to a 1 dimensional evanescent wave  along the spatial $x$ direction  with pulsation $\omega=\sqrt{(\omega_0^2-\kappa^2)}< \omega_0$).\\
\indent The second remark concerns the structure of the far-field and  its description by an inhomogeneous d'Alembert equation. It is clear from the  general structure of  Eq.~\ref{1b} that the  non-linearity of the wave equation can be neglected in the far-field since $|u|\rightarrow 0$ and  therefore we can use  a linearized approximation $D^2u(x)\simeq 0$ far away from the soliton core.  More precisely, as shown in \cite{Drezet2023}, the far-field of the Lane-Emden soliton  with core trajectory $(C)$ $z(\tau):=[t,\mathbf{z}(t)]$ satisfies the equation: 
\begin{eqnarray}
D^2 u(x)=\int_{(C)}g(z(\tau))e^{iS(z(\tau))}\delta^{4}(x-z(\tau))d\tau\nonumber\\=g(t,\mathbf{z}(t))\delta^{3}(\mathbf{x}-\mathbf{z}(t))e^{iS(t,\mathbf{z}(t))}\sqrt{1-\mathbf{v}^2(t)}.\label{debroglieFF}
\end{eqnarray}
The second line is written in the laboratory frame where at a time $t$ the soliton core behaves as a point-like particle with position $\mathbf{z}(t)$ and velocity $\mathbf{v}(t)=\frac{d}{dt}\mathbf{z}(t)$ (here we assume the trajectory being time-like)\\
\indent The general solution of Eq.~\ref{debroglieFF} reads   
\begin{eqnarray}
 u(x)=u_{\textrm{free}}(x)+\int_{(C)}K(x,z(\tau))g(z(\tau))e^{iS(z(\tau))}d\tau=u_{\textrm{free}}(x)+ u_{\textrm{source}}(x)\label{debroglieFFsolu}
\end{eqnarray} 
where $u_{\textrm{free}}(x)$ is a solution of the homogeneous equation $D^2u(x)=0$, and the propagator $K(x,x')$ satisfies $D^2 K(x,x')=\delta^{4}(x-x')$. 
As it is well known, the choice of the propagator is not univocal only the total field $u(x)=u_{\textrm{free}}(x)+ u_{\textrm{source}}(x)$ has a physical unambiguous and absolute meaning. That means, that we can have different representations of the field given by Eq.~\ref{debroglieFFsolu} by changing of propagator. The most common propagators are the retarded (respectively advanced) one $K_{\textrm{ret}}(x,x')$  (respectively $K_{\textrm{adv}}(x,x')$) associated with radiation  in the forward (respectively in the backward) light-cone with apex at  point $x'$. Advanced waves correspond to anticausal,  features associated with `conspiratorial' absorptions by sources (i.e., violating the second law of thermodynamics) and are therefore often not used. Moreover, this is actually related to the boundary conditions that are the most adapted in order to analyze the specifical problem considered. Indeed, we can always write an arbitrary solution  of  Eq.~\ref{debroglieFFsolu} equivalently as  $u(x)=u_{\textrm{in}}(x)+ u_{\textrm{ret}}(x)$, or $u(x)=u_{\textrm{out}}(x)+ u_{\textrm{adv}}(x)$ and the different free fields are thus not independent. As an example: A typical (in the thermodynamical sense) causal field is generally written  after Sommerfeld as a pure radiative field with boundary conditions $u_{\textrm{in}}(x)=0$. This in turn leads to the free `outcoming' field $u_{\textrm{out}}(x)=u_{\textrm{ret}}(x)-u_{\textrm{adv}}(x)$ which is indeed non singular.\\
\indent As explained in the previous subsection \ref{sub31} in 1925 de Broglie  \cite{debroglie1925,debroglie1925b} considered a time-symmetric field in analogy with works in classical time-symmetric electrodynamics~\cite{Tetrode,Page,Fokker,Wheeler}. Therefore, it seems natural to use such time-symmetric description in connection with Eq.~\ref{debroglieFFsolu}. We first write $u(x)=u_{\textrm{I}}(x)+ u_{\textrm{sym}}(x)$ with $u_{\textrm{I}}(x):=\frac{u_{\textrm{in}}(x)+u_{\textrm{out}}(x)}{2}$ a free wave solution of the homogeneous equation and the time-symmetric source term $u_{\textrm{sym}}(x):=\frac{u_{\textrm{ret}}(x)+ u_{\textrm{adv}}(x)}{2}$. We then assume the boundary condition $u_{\textrm{I}}(x)=0$ (i.e., $u_{\textrm{in}}(x)=-u_{\textrm{out}}(x)=\frac{u_{\textrm{adv}}(x)-u_{\textrm{ret}}(x)}{2}$). The $u$-field of Eq.~\ref{debroglieFFsolu} reads now :
\begin{eqnarray}
 u(x)=\int_{(C)}K_{\textrm{sym}}(x,z(\tau))g(z(\tau))e^{iS(z(\tau))}d\tau \label{debroglieFFsoluSym}
\end{eqnarray} with $K_{\textrm{sym}}(x,x')=\frac{K_{\textrm{ret}}(x,x')+K_{\textrm{adv}}(x,x')}{2}$ the time-symmetric propagator. In absence of external field (i.e., $A_\mu(x)=0$) this propagator reads: \begin{eqnarray}
K^{(0)}_{\textrm{sym}}(x,x')=\frac{\delta[(x-x')^2)]}{4\pi}=\frac{1}{2}[\frac{\delta(t-t'-R)}{4\pi R}+\frac{\delta(t-t'+R)}{4\pi R}]\label{green0}
\end{eqnarray} (with $R=|\mathbf{x}-\mathbf{x}'|^2$). In presence of an external electromagnetic field $A_\mu(x)$ the propagator reads \begin{eqnarray}
K_{\textrm{sym}}(x,x')=K^{(0)}_{\textrm{sym}}(x,x')+K^{(ref)}_{\textrm{sym}}(x,x')\label{Greentotal}
\end{eqnarray} where  we introduced $K^{(ref)}_{\textrm{sym}}(x,x')=\int d^4yK^{(0)}_{sym.}(x,y)\hat{\mathcal{O}}_y K_{sym}(y,x')$ (using  the operator $\hat{\mathcal{O}}_y:=e^2A(y)^2-ie\partial_y A(y)-2ieA(y)\partial_y$) which defines the reflected part of the propagator resulting from the interaction of the vacuum solution $K^{(0)}_{sym}$ with the potential $A$. Therefore, the $u-$field splits as \begin{eqnarray}
u(x)=u^{(0)}(x)+u^{(ref)}(x).
\end{eqnarray}
\indent Inserting Eq.~\ref{green0} into Eq.~\ref{debroglieFFsoluSym} leads to \cite{Drezet2023}:
\begin{eqnarray}
 u^{(0)}(x)=\frac{1}{2}\left(\left[\frac{g(\tau)e^{iS(z(\tau))}}{4\pi\rho(\tau)}\right]_{\tau_{ret}}+\left[\frac{g(\tau)e^{iS(z(\tau))}}{4\pi\rho(\tau)}\right]_{\tau_{adv}}\right)\label{Lienard}
 \end{eqnarray} with $\rho(\tau)=|(x-z(\tau))\cdot \dot{z}(\tau)|$  and where the retarded proper time $\tau_{ret}$ (respectively advanced proper time $\tau_{adv}$) corresponds to the point $z(\tau_{ret})$  (respectively  $z(\tau_{adv})$) belonging to the trajectory $(C)$ which $u$-radiation propagating along the forward light-cone (respectively backward light-cone) is reaching the point $x$. This $u$-field is clearly reminiscent from the retarded and advanced Lienard-Wiechert potentials in classical electrodynamics and has several remarkable properties.
Most importantly, near the singularity  $x\sim z(\tau)$, i.e.,  for points located at a distance $r=\sqrt{-\xi^2}$ from the singularity in the space-like  (rest-frame) hyperplane $\Sigma(\tau)$, we have approximately \cite{Drezet2023}: \begin{eqnarray}
u^{(0)}(x)=\frac{g(\tau)e^{iS(z(\tau))}}{4\pi r}[1+\frac{\xi\ddot{z}}{2}+\frac{r^2}{2}(i\ddot{S}-(\dot{S}-i\frac{\dot{g}}{g})^2)+\frac{r^2}{2}\frac{d^2}{d\tau^2}\ln{(g)}\nonumber\\+\frac{3}{8}(\xi\ddot{z})^2+\frac{5}{24}(r\ddot{z})^2+O(r^3)].\label{asymp}
\end{eqnarray} From this we deduce at the lowest order $u(x)\simeq\frac{g(\tau)e^{iS(z(\tau))}}{4\pi r}$  and we recover the asymptotic soliton near-field  (i.e., for $r\gg l_0$ but still in the near-field $r\ll \omega_0^{-1}$) discussed in the subsection \ref{sub31}.  Furthermore, for an uniform motion with $\ddot{z}=0$, $\ddot{S}=0$, $\dot{S}=-\omega_0$, $\dot{g}=0$, $\ddot{g}=0$ Eq.~\ref{asymp} leads to $u^{(0)}(x)=\frac{ge^{-i\omega_0\tau}}{4\pi r}[1-\frac{\omega_0r^2}{2}+O(r^3)]\simeq\frac{ge^{-i\omega_0\tau}\cos{(\omega_0 r)}}{4\pi r}$ that is the field associated with the de Broglie stationnary monopole discussed above. Therefore, the time-symmetric far-field matches the near-field.\\
\indent An important feature of this theory concerns the phase of the singular far-field when $r$ tends to zero. Indeed, from Eq.~\ref{asymp} we deduce in absence of external field: \begin{eqnarray}
\frac{u(x)}{u^\ast(x)}=e^{i2\varphi(x)}=e^{i2S}[1+ir^2(\ddot{S}+2\frac{\dot{S}\dot{g}}{g})+O(r^3)]
\end{eqnarray} which can be compared with the Taylor expansion $e^{i2\varphi(x)}=e^{i2\varphi(z)}[1+i2\xi\partial\varphi(z)+O(r^2)]$ and shows that the first-order term vanishes: 
 \begin{eqnarray}
\xi\cdot\partial\varphi(z)=0.
\end{eqnarray}  Yet, by definition $\xi\dot{z}=0$  in  $\Sigma(\tau)$ and consequently  $\dot{z}(\tau)$ is parallel (i.e., proportional) to $\partial \varphi (z)$. In other words, since $\dot{z}^2=1$, we recover, in absence of external field $A(x)$, the  DS guidance formula Eq.~\ref{DSguidance} $\dot{z}(\tau)=-\frac{\partial\varphi(z(\tau))}{\sqrt{(\partial\varphi(z(\tau)))^2}}$. This results is robust and in \cite{Drezet2023} we showed that it survives in presence of an external electromagnetic field when the full propagator Eq.~\ref{Greentotal} must be considered near the singularity. More precisely, we then have $K_{\textrm{sym}}(x,x')\simeq K^{(0)}_{\textrm{sym}}(x,x')e^{-ieA(z)(x-x')}$  and we deduce \begin{eqnarray}
\frac{u(x)}{u^\ast(x)}=e^{i2S}[1-i2e\xi A(z(\tau))+O(r^2)]=e^{i2\varphi(z)}[1+i2\xi\partial\varphi(z)+O(r^2)],\label{ratio}
\end{eqnarray} implying
\begin{eqnarray}
\xi\cdot (\partial\varphi(z)+eA(z))=0.\label{guido1}
\end{eqnarray} Therefore, as before we recover the guidance formula, i.e.,
 \begin{eqnarray}
\dot{z}(\tau)=-\frac{\partial\varphi(z(\tau))+eA(z(\tau))}{\sqrt{(\partial\varphi(z(\tau))+eA(z(\tau)))^2}}.\label{guido2}
\end{eqnarray} 
Some remarks are important concerning this formula:\\
\indent   First, recovering the guidance formula was expected since de Broglie already gave a general  derivation for the $u-$field in 1926-27\cite{debroglie1927}. As reviewed in \cite{DrezetReview} de Broglie deduction is based on the conservation law Eq.~\ref{2d} written as  $[\partial_t +\mathbf{v}_u\cdot\boldsymbol{\nabla}]\log{\rho_u}=-\boldsymbol{\nabla}\cdot \mathbf{v}_u$ where $\textbf{v}_u(x)=-\frac{\boldsymbol{\nabla}\varphi(x)-e\textbf{A}(x)}{\partial_t \varphi(x)+eV(x)}$ is the 3-velocity of the $u-$fluid and $\rho_u(x)=-2f^2(\partial_t \varphi(x)+eV(x))$. Near the singularity de Broglie assumed  $\rho_u(x):=\rho_u(t,\mathbf{x})\simeq F(t,\mathbf{x})/|\mathbf{x}-\mathbf{z}(t)|^2$, where $F$ is a regular function and $\mathbf{z}(t)$ is the singularity trajectory. De Broglie assumption is indeed satisfied by our singular field $u(x)=u^{(0)}(x)+u^{(ref)}(x)\simeq u^{(0)}(x)$ in a reference frame where the singularity is practically at rest (i.e., $|\frac{d}{dt}\mathbf{z}(t)|\ll 1$). From the property $[\partial_t +\frac{d}{dt}\mathbf{z}(t)\cdot\boldsymbol{\nabla}]|\mathbf{x}-\mathbf{z}(t)|=0$ we thus deduce:
\begin{eqnarray}
(\textbf{v}_u(t,\mathbf{x})-\frac{d}{dt}\mathbf{z}(t))\cdot \frac{(\mathbf{x}-\mathbf{z}(t))}{|\mathbf{x}-\mathbf{z}(t)|}=\frac{|\mathbf{x}-\mathbf{z}(t)|}{2}[\boldsymbol{\nabla}\cdot \mathbf{v}_u+[\partial_t +\mathbf{v}_u\cdot\boldsymbol{\nabla}]\log{F}]=O(|\mathbf{x}-\mathbf{z}(t)|)
\end{eqnarray} 
which implies (near the singularity) the guidance formula
\begin{eqnarray}
\textbf{v}_u(t,\mathbf{x})-\frac{d}{dt}\mathbf{z}(t)=0
\end{eqnarray} Using covariant relativistic notations this is mathematically equivalent to Eqs.~\ref{guido1},\ref{guido2}.\\
\indent  Moreover,  and this constitutes our second remark, the guidance formula is not obvious to satisfy since it requires a well defined gradient $\partial\varphi(x)$ for points near the singularity. In our theory based on a time-symmetric contruction this is automatically fulfilled but this would not be the case if we instead assumed  a pure retarded $u_{\textrm{ret}} $ or advanced  $u_{\textrm{adv}}$ field. Indeed, instead of Eq.\ref{ratio} we obtain:
 \begin{eqnarray}
\frac{u_{\textrm{ret/adv}}(x)}{u_{\textrm{ret/adv}}^\ast(x)}=e^{i2S}[1-i2e\xi A(z(\tau)) \mp 2i r\dot{S}+O(r^2)]=e^{i2\varphi(z)}[1+i2\xi\partial\varphi(z)+O(r^2)],\label{ratiob}
\end{eqnarray}  with the minus (respectively plus) sign for a pure retarded (respectively advanced) wave.   We deduce $\xi\cdot (\partial\varphi(z)+eA(z))=\mp r\dot{S}$ which implies a phase discontinuity on the singularity. Watched in the instantaneaous rest frame  $\Sigma(\tau)$ this condition reads in absence of external field:
\begin{eqnarray}
\partial_rS=\hat{\mathbf{r}}\cdot\boldsymbol{\nabla}S=\mp\dot{S}:=\pm \omega_0.
\end{eqnarray}which is clearly reminiscent of the retarded (respectively advanced) singular field
\begin{eqnarray}
u_{\textrm{ret/adv}}(t,r)\sim e^{-i\omega_0 t}\frac{e^{\pm i \omega_0 r}}{r}.
\end{eqnarray}
In other words, only the time-symmetric monopole $u_{sym}(t,r)\sim e^{-i\omega_0 t}\frac{\cos{(\omega_0 r)}}{r}$ removes the phase discontinuity associated with retarded or advanced waves resulting from an unadapted selection of the propagator $K_{\textrm{ret/adv}}(x,z)$.\\
\indent Regarding the guidance formula derived here from the far-field we stress that it apriori only concerns  the $u-$field containing a singularity or a soliton: Not the the $\Psi-$field which is  much more regular and smooth and has a statistical interpretation, i.e., like in the dBB pilot-wave theory.   However, as discussed in subsection \ref{sub31}, the soliton near-field  used in our DS theory requires the phase harmony condition Eq.~\ref{phaseharmony} near the core of the particle, and this imposes a first-order contact between the two phase functions $\varphi$ and $S$ of the $u$ and $\Psi$ fields: $\varphi(z(\tau))=S(z(\tau))$, $\partial_x \varphi(x=z(\tau))=\partial_z S(z(\tau))$ along the curve $z(\tau)$. This idea is of course applicable to the singular field considered in this subsection.  Therefore, we can here also assume a first order contact $\varphi(z(\tau))=S(z(\tau))$, $\partial_x \varphi(x=z(\tau))=\partial_z S(z(\tau))$ and obtain the full guidance formula needed in the DS theory:
\begin{eqnarray}
\dot{z}(\tau)=-\frac{\partial\varphi(z(\tau))+eA(z(\tau))}{\sqrt{(\partial\varphi(z(\tau))+eA(z(\tau)))^2}}=-\frac{\partial S(z(\tau))+eA(z(\tau))}{\sqrt{(\partial S(z(\tau))+eA(z(\tau)))^2}}. \label{guido3}
\end{eqnarray} 
In other words using a first-order contact (and not a second-order contact, i.e.  imposing also $\partial_\mu\partial_\nu \varphi(z(\tau))=\partial_\mu\partial_\nu S(z(\tau))$ as originally assumed by de Broglie and Vigier \cite{debroglie1927,debroglie1956,Vigierthesis}) allows us to develop a self-consistent DS model in both the near-field and far-field.\\
\subsection{`Justifying' the wave equation for the $\Psi-$guiding field}\label{sub33}
\indent The present  theory for the $u-$field left apriori unconstrained or undetermined i) the precise form of the wave equation   for $\Psi$, and  (ii) the physical nature of this $\Psi-$field. First consider point (i): the mathematical form of the wave equation  for the $\Psi-$field. In our theory \cite{Drezet2023} we assumed that the $\Psi-$field obeys the linear Klein-Gordon  equation. This choice can be approximately justified.\\ 
\indent  For this purpose we start with the phase harmony relation Eq.~\ref{phaseharmony} defining the phase $\varphi(x)$ in the vicinity of the soliton-core trajectory  $z(\tau)$ characterized by the phase $\varphi(z(\tau):=S(z(\tau))$. We dont have here to assume that the soliton is coupled to an external physical  wave $\Psi$ with phase $S$ guiding the soliton. Indeed, it is enough to show that from $\varphi(z(\tau):=S(z(\tau))$ obtained from our wave equation for $u$ we can construct a wave equation  for a $\Psi-$field having the properties of the linear Klein-Gordon equation.  The method goes back to Vigier and R\'egnier \cite{VigierCR,RegnierCR,Vigierthesis}. We assume that the soliton trajectory $z(\tau)$ exists and the phase is given in its vicinity by the phase-harmony condition Eq.~\ref{phaseharmony}. Now, we consider a statistical ensemble of solitons and impose a local conservation law 
\begin{eqnarray}
\partial_z[-2\mu(z)(\partial_z S(z)+eA(z))]=0,\label{2bb}
\end{eqnarray}  where $\mu(z)\geq 0$ plays the role of a density  for the statistical fluid in the configuration space of the particle with path $z(\tau)$. We thus define a wave field given by \begin{eqnarray}
\Psi(z):=\sqrt{\mu(z)}e^{iS(z)}\end{eqnarray} and this allows us to rewrite Eq.~\ref{2bb} as 
\begin{eqnarray}
i\partial_z[\Psi^\ast(z)D_z\Psi(z)-\Psi(z)D_z^\ast\Psi^\ast(z)]=0,\label{2bc}
\end{eqnarray}with $D_z:=\partial_z+ieA(z)$. After transformation Eq.~\ref{2bc} reads 
\begin{eqnarray}
i[\Psi^\ast(z)D^2_z\Psi(z)-\Psi(z)(D_z\Psi(z))^\ast]=0.\label{2bd}
\end{eqnarray} or equivalently $\textrm{Imag}[\Psi^\ast(z)D^2_z\Psi(z)]=0$. Now, the real part $\textrm{Real}[\Psi^\ast(z)D^2_z\Psi(z)]$ is left unconstrained by this procedure and this leads to the general condition 
\begin{eqnarray}
\Box_z a(z)-(\partial_z S(z)+eA(z))^2a(z)=-W(z)\label{prewaveeq}
\end{eqnarray} 
with $W(z)\in \mathbb{R}$ a function of $z$, and finally implies:
\begin{eqnarray}
D^2_z\Psi(z)=-W(z)\frac{\Psi(z)}{|\Psi(z)|}.\label{waveeq}
\end{eqnarray} Clearly, if we separate the real and imaginary part  in  the wave equation Eq.\ref{waveeq} we recover Eq.~\ref{2bb} and obtain the generalized Hamilton-Jacobi equation
\begin{eqnarray}
(\partial_z S(z)+eA(z))^2=\frac{W(z)}{a(z)}+Q_\Psi(z)\label{treu}
\end{eqnarray} with $Q_\Psi(z)=\frac{\Box_z a(z)}{a(z)}$ the quantum potential.   In order to close our `derivation' of the wave equation   we now consider the semiclassical regime where $Q_\Psi(z)=\frac{\Box_z a(z)}{a(z)}\ll 1$ and this implies $(\partial_z S(z)+eA(z))^2\simeq W(z)/a(z)$. In order to recover classical physics for a particle of mass $\omega_0$ in an external field $A(z)$ we impose $W(z):=\omega_0^2a(z)$ and this leads to the linear Klein-Gordon equation
\begin{eqnarray}
D^2_z\Psi(z)=-\omega_0^2\Psi(z).\label{waveeqf}
\end{eqnarray} 
We can also justify the choice $W(z):=\omega_0^2a(z)$ on physical ground: If we assume that in the remote past the soliton has a uniform inertial motion defining in the rest-frame a pulsation $\omega_0$ we can identify this motion with $Q_\Psi=0$ in Eq.\ref{treu}. Consequently, in order to fulfill the first-order contact hypothesis we  impose $W(z):=\omega_0^2a(z)$.\\
\indent  The previous reasoning motivates the choice for a guiding field obeying a linear Klein-Gordon equation  but clearly doesnt impose it.  This shows that the choice of a guiding $\Psi-$field is not here dicted  only by physical reasonings but also by practical features associated with the simplicity of a linear wave equation compared to a non-linear one.\\
\indent This in turn  leads us to point (ii): What is the physical meaning of the $\Psi-$field. Indeed, since we are here proposing a minimal model with only a single  fundamental $u-$field,  the $\Psi-$field cannot be a fundamental independent field interacting with the $u-$field (such a different approach has beeen developped by us in \cite{Drezet2023b}).   Here, we suggest  to interptet the $\Psi-$field as the natural extension of the action $S(z)$ introduced in the old Hamilton-Jacobi equation for a point-like particle.   Here, the wave  $\Psi(z):=\sqrt{\mu(z)}e^{iS(z)}$ is seen as a mathematical tool for describing the motion of the soliton with trajectory $z(\tau)$ and is clearly similar to the role played by $S(z(\tau))$ obeing  $(\partial_z S(z)+eA(z))^2=\omega_0^2$ in the old classical theory.  Moreover,  in the classical Hamilton-Jacobi  theory the 3D configuration space, with vectors  $\mathbf{z} \in \mathbb{R}^3$, defines the set of all possible positions and trajectories for the point-like particle and must be distinguished from the 3D physical space with vector  $\mathbf{x}\in \mathbb{R}^3$    where the extended soliton is evolving.  In the DS theory developed here the complex wave function $\Psi(z):=\sqrt{\mu(z)}e^{iS(z)}$ defines a generalization of the Hamilton-Jacobi function adapted to the dynamics of a soliton and $z$ defines a configuration space for the center of the soliton.\\ 
\indent At the philosophical level we stress that there is an old debate  between advocates of the dBB pilot-wave or `Bohmian' mechanics concerning the  physical status of the wave function $\Psi$. De Broglie always emphasized that for him the wave function $\Psi$ is not a fundamental of `objective' field  but instead a `subjective' probabilistic field \cite{debroglie1956}. Quite similarly,  D\"{u}rr, Goldstein and Zangh\`{i} \cite{Durr}  wrote that the wave function $\Psi$ is not a physical external agent acting on the particle but better a mathematical (nomological) object used for describing the quantum law of particle (similar to the Hamiltonian function $H(q,p,t)$ in classical mechanics). A physical point clearly in favor of this view is the absence of retroaction of the particle on the $\Psi-$wave in the dBB theory. An objection often made against this nomological view is that in classical mechanics the Hamiltonian function is given whereas the wave function  depends on the choice of initial conditions.  However, we can easily counter the objection:  The problem is not actually the nomological view itself but instead the comparison with the Hamiltonian.   As we saw, and in agreement with  de Broglie, the good comparison concerns $\Psi$ and $S$, i.e., the wave function  and the action in the old Hamilton-Jacobi theory. Indeed, $\Psi$ and $S$ are both dependent on initial conditions, are defined in the configuration space, can evolve in time, and are used to classify ensembles of possible particle trajectories $z$. However, we can also perhaps justify the psychological resistance against the nomological view by the non-intuitive features observed in the dBB theory. For example, as already pointed out by de Broglie in 1930 \cite{debroglie1930} concerning interference in the double-hole  experiment, it looks as if the dynamical motion of the single particle  (going through one hole) is affected by alternative motions (going through the second hole) which didn't occur but were potentially possible. In classical physics this doesn't happen and possible trajectories coming from each hole are just crossing each other, i.e., the particles going through one hole are completely unaffected by the presence of the second hole not crossed. In the DS approach this non intuitive aspect of the pilot-wave dynamics is explained by the existence of the $u-$field associated with an extended physical phenomena surrounding the soliton core and involving the whole environement of the wave. In other words,  the $\Psi-$dynamics is just an effective description of the soliton core motion $z(\tau)$ that in practice is neglecting an important   part of the $u-$wave propagation associated with the retarded and advanced contributions focused on the soliton.
\section{The many-body problem}\label{sec4}
\begin{figure}[h]
\includegraphics[width=8cm]{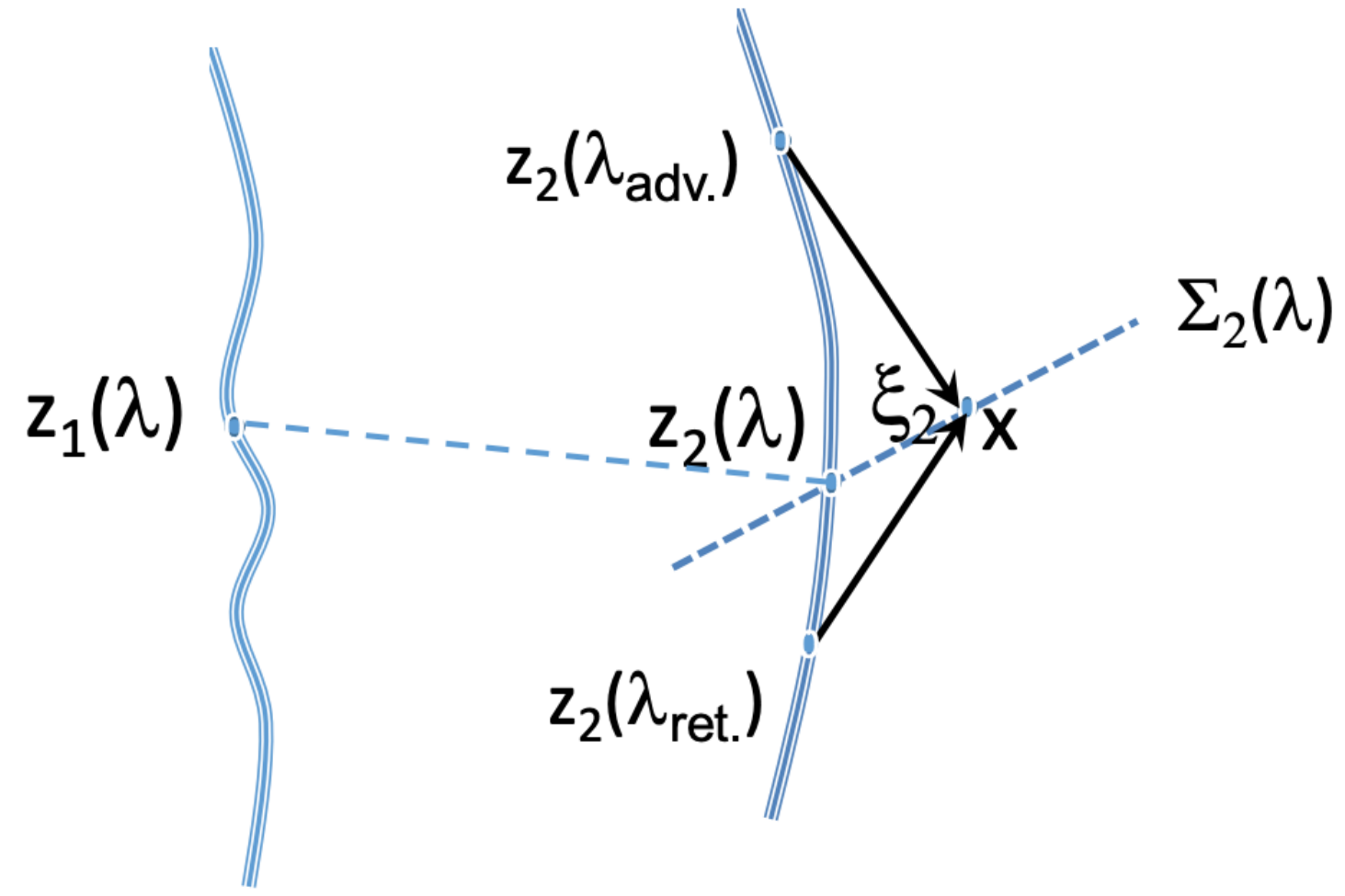} 
\caption{Determination of the space-like hyperplane  $\Sigma_2(\lambda)$ for a system of two particles 1 and 2 with synchronized trajectories 
 $z_1(\lambda)$, $z_2(\lambda)$. For a point $x$ located near the trajectory $z_2$ the hyperplane $\Sigma_2(\lambda)$ is defined by the relation 
 $\xi_2\cdot\dot{z}_2(\lambda)=0$. Once the corresponding point $z_2(\lambda)$ is unambiguously defined the value of $\lambda$ fixes the entangled position $z_1(\lambda)$ associated with the other particle.  Also shown are retarded and advanced positions of the particle 2 used in the description of the $u-$field.    } \label{Fig1}
\end{figure}

\indent The previous model based on a local but non-linear wave equation can be extended to the case of many non interacting solitons coupled to external electromagnetic fields \cite{Drezet2023}.  The idea is to find a multisoliton solution $u(x)$ of Eq.~\ref{1b}. Moreover, an explicit and analytical formula for describing such a system is probably impossible to find.  Therefore, we instead assume that the far-field of  a given soliton is decaying quickly with the distance to the center. Here, we work in the approximation where the perturbation near the location $z_j$ of the $j^{th}$ soliton $\delta u_i(z_j)$, and associated with the $u-$field created by a different soliton labeled by the index $i$,  is small enough and can be neglected compared to the near-field of the $j^{th}$ soliton itself (i.e., if $|\delta u_i(z_j)|\ll \frac{g_0\sqrt{\alpha_j}}{4\pi l_0}$). Physically this makes sense if the various solitons are separated by distances  $R_{i,j}\gg l_0$ (this will be true in general if the soliton typical extension is very small compared to the Compton wavelength and other characteristic lengths of the system).\\
\indent In the near-field of the  $j^{th}$ soliton with trajectory $z_j$ we can apply the method described in Sec.~\ref{sub31}. In particular, using the phase-harmony condition we can define for points $x$ in the local hyperplane $\Sigma_j(\lambda)$ the phase  $\varphi(x)$ if $x\simeq z_j(\lambda)$:
\begin{eqnarray}
\varphi(x) \simeq S_N(\{z_j(\lambda)\}) -eA(z_j(\lambda))\xi_j+B_j(\{z_j(\lambda)\})\frac{\xi_j^2}{2}+O(\xi_j^3).\label{phaseharmonyMany}
\end{eqnarray}  Like for a single soliton the local hyperplane is defined by the condition $\xi_j\cdot\dot{z}_j(\lambda)=0$ with $\xi_j=x-z_j(\lambda)$ and   $\dot{z}_j(\lambda):=\frac{d}{d\lambda}z_j(\lambda)$ is the soliton center velocity. $B_j(\{z_j(\lambda)\})$ is a collective coordinate measuring the deformation of the $N$ solitons.  In this discussion $\lambda$ is a common evolution parameter for the moving points along the various trajectories $z_1(\lambda),...,z_N(\lambda)$. Therefore, $\lambda$ defines a `common time'  for the $N$ synchronized particles. We stress that Eq.~\ref{phaseharmonyMany} concerns points $x$ contained in the local hyperplane $\Sigma_j(\lambda)$ of the $j^{th}$ soliton with trajectory $z_j(\lambda)$. This means that for a fifferent soliton, let says the $k^{th}$, we need an equivalent equation (this explains why $B_j(\{z_j(\lambda)\})$ is labeled by the soliton number $j$ or $k$: $B_k(\{z_j(\lambda)\})$). We have thus $N$ local expansions to consider in this approach, each one corresponding to a different soliton solution of the same nonlinear equation. Mathematically, this means that $\varphi(x):=\varphi_j(x|\{z_j(\lambda)\})$ in Eq.~\ref{phaseharmonyMany}, i.e., that the phase is locally conditionned on the knowledge of the $N$ synchronized trajectories once the hyperplane $\Sigma_j(\lambda)$ for the $j^{th}$ particle  is defined.  This operation is geometrically $x$-dependent and unambiguous  at least for points located not too far from the particle trajectories (see Figure \ref{Fig1}).   \\
\indent  Moreover, nothing has been yet said about the choice of the action function $S_N(\{z_j(\lambda)\}):=S_N(z_1(\lambda),...,z_N(\lambda))$. In the context of the dBB pilot-wave theory it is natural to introduce the wavefunction $\Psi_N(\{x_j\})=a_N(\{x_j\})e^{iS_N(\{x_j\})}$ solution of the set of $N$ coupled Klein-Gordon equations:  
$ D_j^2\Psi_N(\{x_j\})=-\omega_0^2\Psi_N(\{x_j\})$ with $D_j:=\partial_j +ieA(x_j)$ and $\partial_j$ the 4-gradient operator for the $j^{th}$ particle.   Using the polar representation we can  write  
\begin{subequations}
\begin{eqnarray}
(\partial_j S_N(\{x_j\})+eA(x_j))^2=\omega_0^2+Q_{\Psi_N,j}(\{x_j\}):=\mathcal{M}^2_{\Psi_N,j}(\{x_j\}) \label{2N}\\
\partial_j[a_N^2(\{x_j\})(\partial_j S_N(\{x_j\})+eA(x_j))]=0,\label{2bN}
\end{eqnarray}
\end{subequations} with $Q_{\Psi_N,j}(\{x_j\})=\frac{\Box_j a_N(\{x_j\})}{a_N(\{x_j\})}$ a quantum potential.  In the context of the relativistic dBB theory the particle velocity for the $j^{th}$ particle is supposed to be 
\begin{eqnarray}
\frac{dz_j(\lambda)}{d\lambda}=-(\partial_j S_N(\{z_j(\lambda)\})+eA(z_j(\lambda))\sqrt{\left(\frac{\dot{z}_j(\lambda)\dot{z}_j(\lambda)}{\mathcal{M}^2_{\Psi_N,j}(\{z_j(\lambda)\})}\right)}.\label{PWIguidanceGenN}
\end{eqnarray} Once initial conditions  $z_1(0),z_2(0),...,z_N(0)$ are given this set of coupled equations can be integrated to obtain  $N$ coupled (i.e., entangled) trajectories for the $N$ particles. We stress that Eq.~\ref{PWIguidanceGenN} is general and valid whatever the sign of  $\mathcal{M}^2_{\Psi_N,j}$ (in particular  if $\mathcal{M}^2_{\Psi_N,j}<0$ the particle is moving faster than light as explained in the Appendix).\\
\indent An important feature of dBB trajectories obtained here is that they can be used to compute the wavefunction $\Psi_N(\{z_j(\lambda)\})$ knowing $\Psi_N(\{z_j(0)\})$. More precisely we have: 
\begin{subequations}
\begin{eqnarray}
\frac{d}{d\lambda}\ln{a^2_N(\{z_j(\lambda)\})}=-\sum_j\sqrt{\left(\frac{\dot{z}_j(\lambda)\dot{z}_j(\lambda)}{\mathcal{M}^2_{\Psi_N,j}(\{z_j(\lambda)\})}\right)}\partial_j\left(\dot{z}_j(\lambda)\sqrt{\left(\frac{\mathcal{M}^2_{\Psi_N,j}(\{z_j(\lambda)\})}{\dot{z}_j(\lambda)\dot{z}_j(\lambda)}\right)}\right)\label{normN}\\
\frac{d}{d\lambda}S_N(\{z_j(\lambda)\})=-\sum_j\textrm{sign}(\dot{z}_j(\lambda)\dot{z}_j(\lambda))\sqrt{\left(\dot{z}_j(\lambda)\dot{z}_j(\lambda)\mathcal{M}^2_{\Psi_N,j}(\{z_j(\lambda)\})\right)}\nonumber\\ +eA(z_j(\lambda))\cdot\dot{z}_j(\lambda)\label{actionN}
\end{eqnarray}
\end{subequations}
where Eq.~\ref{normN} is deduced from the $N$ conservation rules  $\partial_j\left(a^2_N\dot{z}_j\sqrt{\left(\frac{\mathcal{M}^2_{\Psi_N,j}(\{z_j\})}{\dot{z}_j\dot{z}_j}\right)}\right)=0$ (i.e., Eq.~\ref{2bN}) and the definition $\frac{d}{d\lambda}:=\sum_j\dot{z}_j\partial_j$, and similarly  Eq.~\ref{actionN} from the definition $\frac{d}{d\lambda}S_N=\sum_j\dot{z}_j\partial_jS_N$ (i.e., equivalent to a Lagrangian for the $N$ dBB particles) and Eq.~\ref{PWIguidanceGenN}.\\
\indent Moreover, we point out that contrarily to what occurs in the nonrelativistic regime (i.e., based on the many-body Schr\"odinger equation) the dBB trajectories obtained here from the set of $N$ coupled Klein-Gordon equations is in general not able to reproduce all statistical predictions of standard quantum mechanics for every times (the theory is said to benot statistically transparent). Indeed, we in general apriori don't know how (and we don't know if it is even possible) to combine  the $N$ partial conservation laws $\partial_j\left(a^2_N\dot{z}_j\sqrt{\left(\frac{\mathcal{M}^2_{\Psi_N,j}(\{z_j\})}{\dot{z}_j\dot{z}_j}\right)}\right)=0$ (i.e., Eq.~\ref{2bN}) into a single `master' equation defining a probabilistic conservation law for the $N-$paths. Of course, in the non relativistic regime the situation goes easier since   Eq.~\ref{2bN}  reduces to $\partial_{t_j}a_N^2+\boldsymbol{\nabla}_j(a_N^2\mathbf{v}_j(t_j))=0$. In this nonrelativistic regime we can introduce a single common time $t:=t_1=...=t_N$ such that $\partial_t=\sum_j\partial_{t_j}$ and we deduce
\begin{equation}
\partial_t a_N^2+\sum_j\boldsymbol{\nabla}_j(a_N^2\mathbf{v}_j(t))=0
\end{equation} which recovers the standard Bohmian probability law for the many-body Schr\"odinger equation with the definition $a_N:=a_N(t,\mathbf{z}_1(t),...,\mathbf{z}_N(t))$. Here  $a_N=|\Psi_N|^2$ defines the density of probability in the configuration space in agreement with Born's rule.  However, despite the present limitations it is possible to show (and the mathematical details will not be given here but in a subsequent publication) that the relativistic dBB trajectories given by Eq.~\ref{PWIguidanceGenN} are asymptotically  statistically transparent. This means that such $N$ paths can be used to recover statistical predictions of quantum mechanics in scattering processes where interactions between particles and fields are well localized in space-time and where particles can be considered as initially independent (i.e., unantangled).  With such restrictions the theory is physically satisfying for all practical purposes. In the following we will not consider this problem anymore and accept the physical relevance of Eq.~\ref{PWIguidanceGenN} for founding a self-consistent dBB theory.\\
\indent Going back to our DS theory and to Eq.~\ref{phaseharmonyMany} we now have a set of N synchronized  dBB trajectories $z_j(\lambda)$  used to define local rest frames 	and hyperplanes  $\Sigma_j(\lambda)$. The general method developed in Section \ref{sub31} is thus applicable. In particular, using fluid conservation Eq.~\ref{newdfluid} allows to determine the $N$ deformation coefficients $B_j(\{z_j(\lambda)\})$ obeying to the set of coupled equations  
\begin{eqnarray}
B_j(\{z_j(\lambda)\})=\frac{\dot{z}_j}{2\sqrt{\left(\dot{z}_j\dot{z}_j\right)}}\partial_j \sqrt{\mathcal{M}^2_{\Psi_N,j}(\{z_j\})}
\end{eqnarray} that generalizes for $N$ solitons the results  Eqs.~\ref{inteG} and \ref{inteGSu} deduced for a single soliton. In order to complete the description we need to evaluate the amplitude $f(x)$ of the $u-$field for points such as $x\simeq z_j(\lambda)$ in the hyperplane $\Sigma_j(\lambda)$. Like for $\varphi$ and Eq.~\ref{phaseharmonyMany} we have locally $f(x):=f_j(x|\{z_j(\lambda)\})$ and this amplitude obeys to a nonlinear equation  generalizing Eq.~\ref{ODE}, i.e.
\begin{eqnarray}
\mathcal{M}^2_{\Psi_N,j}(\{z_j(\lambda)\})f_j(x|\{z_j(\lambda)\})+\boldsymbol{\nabla}^2f_j(x|\{z_j(\lambda)\})\simeq -\frac{3l_0^2}{(\frac{g_0}{4\pi})^4}f^5_j(x|\{z_j(\lambda)\}).
\label{ODEmany}\end{eqnarray}
As in Section \ref{sub31} the integration of this equation leads to 
\begin{eqnarray}
f_j(x|\{z_j(\lambda)\}):=F_{\alpha_j}(r_j)=\frac{\sqrt{\alpha_j}g_0}{4\pi}\frac{1}{\sqrt{\alpha_j^2r_j^2+l_0^2}}\label{Lanesolutionscaledmany}
\end{eqnarray}
with $r_j$ the radial distance to the $j^{th}$ soliton center, and where we have: 
\begin{eqnarray}
\alpha_j(\{z_j(\lambda)\})=\alpha(\{z_j(-\infty)\})\sqrt{\left(\frac{|\mathcal{M}_{\Psi_N,j}(\{z_j(\lambda)\})|}{|\mathcal{M}_{\Psi_N,j}(\{z_j(-\infty)\})|}\right)}
=\sqrt{\left(\frac{|\mathcal{M}_{\Psi_N,j}(\{z_j(\lambda)\})|}{\omega_0}\right)}
\end{eqnarray} 
if we assume $\alpha_j(\{z_j(-\infty)\})=1$ and $\mathcal{M}_{\Psi_N,j}(\{z_j(-\infty)\})=\omega_0$ (see Apendix \ref{appendix}). This analysis complete our description of the $N-$solitons near-field.\\
\indent  The description of the $N-$solitons far-field can be done similarly by generalisation of the method developed in Section \ref{sub32}. In the far-field regime the $u-$field obeys to a linear equation except along singular lines corresponding to the $N$ trajectories $z_j(\lambda)$. The field at point $x$ reads $u(x)=\sum_ju_j(x)$ where $u_j(x)$ is a solution of  Eq.\ref{debroglieFF}. Therefore we have:
\begin{eqnarray}
D^2 u(x)=g_0\sum_j\int d\lambda\sqrt{|\dot{z}_j(\lambda)\dot{z}_j(\lambda)|} \frac{e^{iS_N(\{z_j(\lambda)\})}}{\sqrt{\alpha_j(\{z_j(\lambda)\})}}\delta^{4}(x-z_j(\lambda))\label{singularN}
\end{eqnarray} where $\frac{g_0}{\sqrt{\alpha_j(\{z_j(\lambda)\})}}:=g_j(\{z_j(\lambda)\})$ defines a coupling constant for each individual singularity. The solution of Eq.~\ref{singularN} we consider reads:
\begin{eqnarray}
u(x)=g_0\sum_j\int d\lambda\sqrt{|\dot{z}_j(\lambda)\dot{z}_j(\lambda)|} K_{\textrm{sym}}(x,z_j(\lambda))
\frac{e^{iS_N(\{z_j(\lambda)\})}}{\sqrt{\alpha_j(\{z_j(\lambda)\})}}\label{soluN}
\end{eqnarray} with $K_{\textrm{sym}}(x,z_j(\lambda))$ the time-symmetric Green propagator given by Eq.~\ref{Greentotal}.\\
\indent The picture we get  in the far-field is thus the following: \\
\indent (i) Starting from the dBB pilot-wave theory for $N$ relativistic scalar particles we define $N$ generally entangled particles trajectories $z_1(\lambda),...,z_N(\lambda)$ guided by the wavefunction $\Psi_N(\{z_j(\lambda)\})$.\\
\indent (i) To each trajectory $z_j(\lambda)$ we associate a moving singularity term in the linear but inhomogenous equation  Eq.~\ref{singularN}.\\
\indent (iii) The solutions we consider are the  $N$ time-symmetric fields $u_j(x)$ which sum is given by Eq.~\ref{soluN} and which depend on the time-symmtric propagator  
$K_{\textrm{sym}}(x,z_j(\lambda))$.\\ 
\indent The consistency of the whole picture, as explained before for the near-field, relies on the assumption that the $N$ solitons are non-interacting, i.e., that we neglect the effect of soliton $u_j$ on soliton $u_k$ for any pair $j,k$. Relaxing this condition could for example imply that we take into account the electromagnetic interaction between solitons and this would ultimately require a development of quantum electrodynamics for solitons (with particle and antiparticle creation). A second possibility for extending the theory could be to include the interaction of solitons when  we can not neglect the perturbation $\delta u_i(z_j)$  compared to $u_i(z_i)$. In this regime new effects could potentilally appear going beyond the usual predictions of quantum mechanics (i.e. beyond the guidance formula of the dBB theory). This clearly opens interesting perspectives for futur works.\\
\indent An important feature of this DS approach is that we started from a local but non-linear equation for the  $u-$field and nevertheless we were able to find solitonic solutions  driven by a phase $S_N(\{z_j(\lambda)\})$ which in general implies nonlocal action at a distance between the (dBB) trajectories. Is that not a contradiction after all? As we analyzed in \cite{Drezet2023} the fundamental aspect of this theory is the time-symmetry associated with the half sum of advanced and retarded waves in the far-field. As we will now discuss  this explains how to remove contradictions and even to justify and explain the dBB nonlocality as an effective feature of our nonlinear dynamics involving time-symmetric fields.     
\section{Discussion: Superdeterminism and effective nonlocality \`a la Bohm}\label{sec5}    
\begin{figure}[h]
\includegraphics[width=13cm]{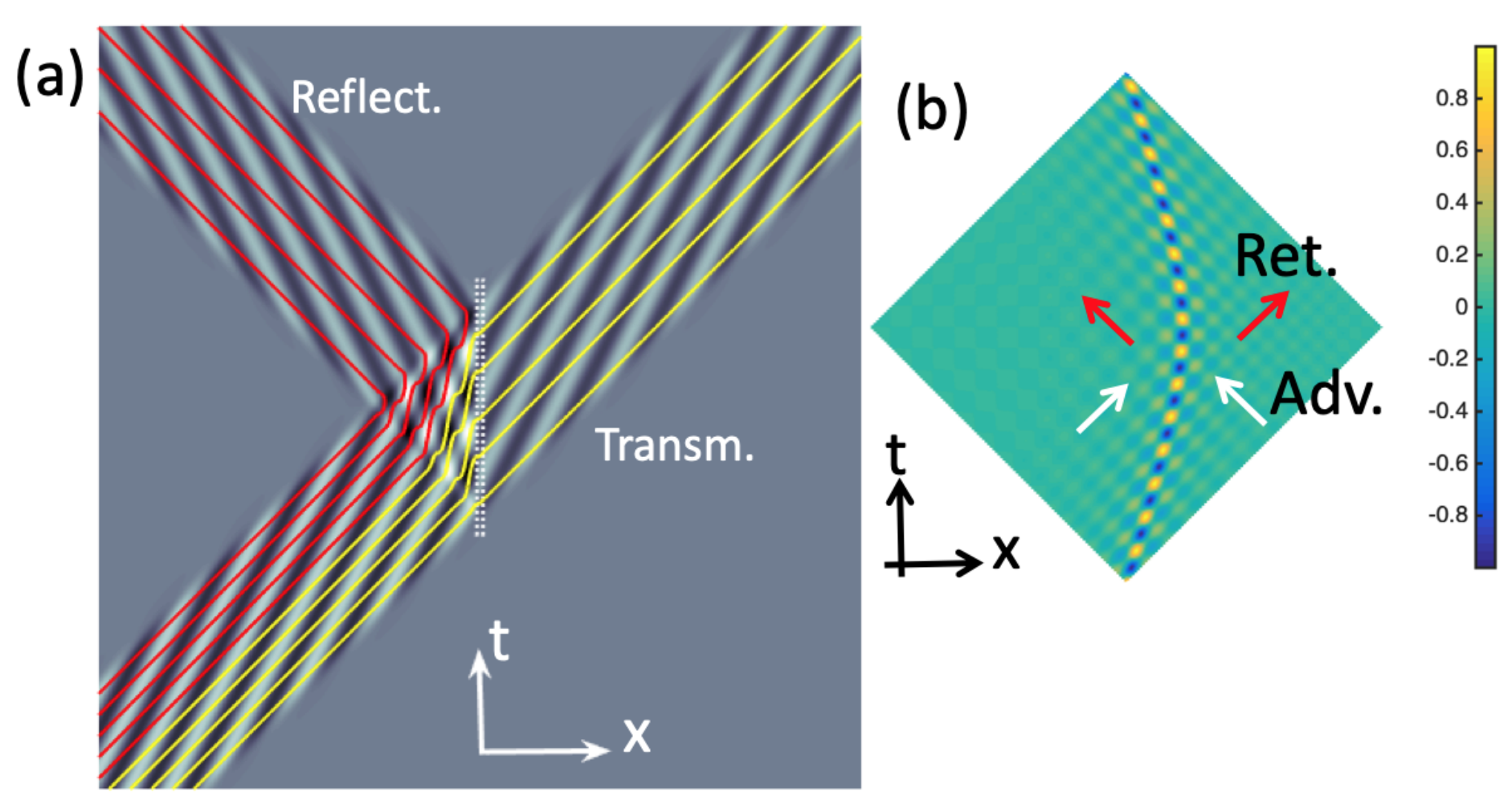} 
\caption{(a) A typical one dimensionnal scattering experiment with a $\Psi$ Schr\"odinger wave-packet impinging on a 50/50 beam splitter (strongly localized field) and the dBB particle trajectories associated (details in \cite{DrezetEntropy}). Half of the trajectories are reflected (red curves) and half transmitted (yellow curves). (b) A typical $u$ time-symmetric field guided by an idealized  reflected dBB trajectory of  (a): See the model in \cite{Drezet2023} for more details). The arrows indicate the presence of retarded and advanced components   propagating forward and backward in time from any points of the dBB main trajectory.} \label{Fig2}
\end{figure}

\indent The present DS theory, with its underlying time-symmetry, has some remarkable consequences for discussing  the nature of causality in quantum mechanics. 
Indeed, in the standard dBB pilot-wave theory the first-order guidance formula  $\dot Q(t)=F_\Psi(Q(t),t)$  for the set of coordinates (local beables) $Q(t):=[q_1(t),q_2(t),...]$ at time $t$ leads to trajectories $Q(t)=G_\Psi(t,Q(t_{in}),t_{in})$ requiring the knowledge  ot the initial positions  $Q(t_{in})$, a time $t_{in}$. As it is has been often emphasized this Bohmian evolution is strongly contextual and also  presupposes a preferred space-time foliation. Nevertheless, this dBB theory preserves some natural features present in the old classical dynamics: Namely  the knowledge of the past state is needed and sufficient to predict the future evolution of the system. In this Cauchy problem, the integration of the guidance formula, i.e  for the Schr\"odinger  or Klein-Gordon equations, is thus naturally obtained once we know the physical state defined along a space-like hypersurface $\Sigma(t_{in})$ located in the past. In our DS theory we preserved the validity of the dBB theory but the trajectories are used to guide solitons having time-symmetric profiles in space-time. Indeed because of the presence of the time-symmetric Green propagator $K_{\textrm{sym}}(x,z_j(\lambda))$ in Eq.~\ref{soluN} the moving solitons (i.e., moving singularities in the far-field approximation) emit natural retarded  waves $u_{\textrm{ret}}(x)$ into the future time direction, but  also more exotic advanced waves $u_{\textrm{adv}}(x)$ `propagating' into the past direction.                 The direct consequence is that any space-like hypersurface like  $\Sigma(t_{in})$  contains informations about physical interactions acting upon the particle in the future. It is not difficult to see that this information coming from the future and affecting the initial state can be interpreted as a form of superdeterminism associated with the retrocausal waves emitted by the particles in the future.\\
\indent As an illustration, consider  the case of a dBB particle interacting with a 50/50 beam-splitter  as studied for example in \cite{DrezetEntropy} (see Figure \ref{Fig2} (a)).          
As explained in \cite{DrezetEntropy} the $\Psi-$wave packet associated with an incident particle (represented by a quasi monochromative wave in Figure \ref{Fig2} (a)) is impiging  on the beam splitter represented by an external field strongly localized in space and also potentially in time.   The system is tuned in order to have 50\% of the particles reflected and 50\% transmitted. In the dBB pilot-wave theory this implies that half of the initial possible trajectories will be reflected or transmitted and that the  exact outcome of the experiment depends precisely on the initial condition (i.e., position) of the particle in the incident beam. Since this initial coordinate is unknown, i.e., hidden, to the observer the result is described by probability (see \cite{DrezetEntropy} for a discussion) in agrement with Born's rule of standard quantum mechanics. Moreover, in the dBB theory one sometimes question the role of the empty channel not chosen by the particle. Suppose for instance that the particle is reflected then a $\Psi-$wave guides the particle in the reflected branch   but an `empty wave' should apriori  propagates in the non occupied channel or branch.  The existence of this empty wave is apriori inferred from the fact that  we could by adding mirrors and a second beam splitter in the paths of  the waves create an interferometer where the influence of the wave propagating in the empty channel is required in order to recover the observed results. Already in 1930 \cite{debroglie1930} de Broglie considered this issue as problematic for the pilot-wave theory since the empty wave should carry energy and this has never directly been detected despite many attempts \cite{Selleri}. Of course we can use a nomological picture and refuse to attribute any physical content to the empty wave. However, the problem apriori survives in the DS theory  developed by de Broglie in the 1950's where the soliton was expected to loose (i.e., radiate) progressively its energy after interacting with several beam splitters \cite{debroglie1956,Selleri}. In our DS time-symmetric approach the situation is clearly different because of the time-symmetry involved.  Consider for example a typical reflected dBB trajectory from Figure \ref{Fig2} (a) i.e., associated with the motion $z(\lambda)$ of the soliton core. the $u-$field of such a soliton is computed in Figure \ref{Fig2} (b) for a simplified model. In agreement with Eq.~\ref{debroglieFFsoluSym} involving $K_{\textrm{sym}}(x,z(\tau))$ this $u-$field is the half sum  of a retarded and advanced contributions $\frac{u_{ret}(x)+u_{adv}(x)}{2}$.  Furthermore the advanced field $u_{adv}(x)$ `propagates' backward in time and this  even before the particle crossed the beam-splitter. Therefore, in the remote past, i.e., before the interaction  with the beam splitter occured,  there is an advanced field $u_{adv}(x)$ which  will focus on the particle singularity at a later time, i.e., after the interaction with the beam splitter. This advanced field carries an information  and energy on the later interaction: Something which is looking retrocausal of conspiratorial. Indeed if we watch the time evolution normally, i.e., from past to future, what we see is a perfectly well tuned $u-$field converging on the particle and ariving precisely at the good moment in order to fulfill the wave equation. From the point of view of normal causality going from past to future   this is a form of superdeterminism.             
\begin{figure}[h]
\includegraphics[width=9cm]{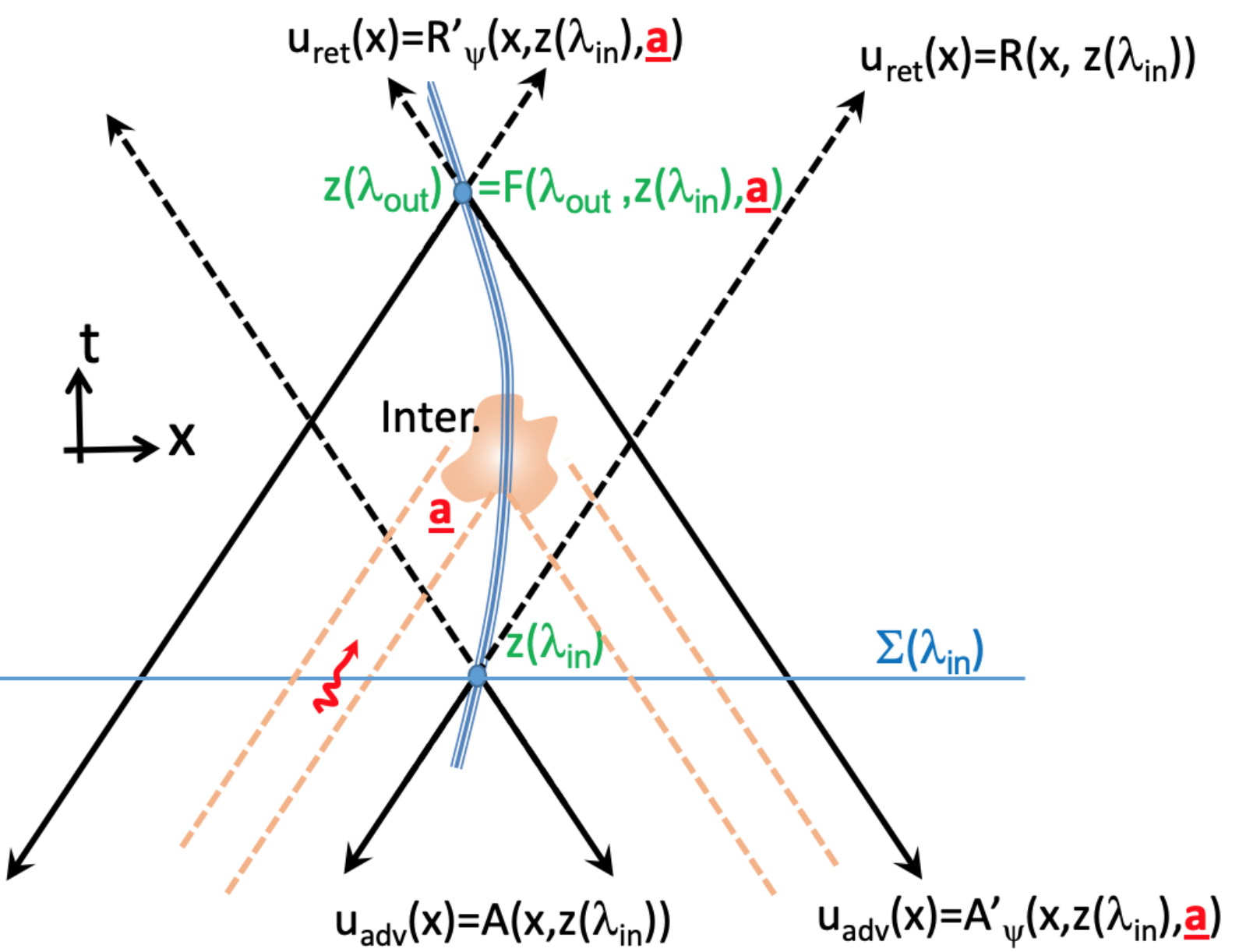} 
\caption{A generic dBB trajectory passing through an interaction zone where a classical field is characterized by the external settings $\underline{\bold{a}}$ (the settings can be modified within the past ligth cones (orange color) with appexes in the interaction volume). As explained in the main text the dBB motion after the interaction  will depend on the initial condition  $z(\lambda_{in})$ and the settings $\underline{\bold{a}}$. Moreover the retarded  (respectively advanced) $u-$waves emitted by the particle propagate the information about the information very far away from the interaction zone. In particular, the advanced waves $u_{adv}(x)$ send information about  $\underline{\bold{a}}$ at points located  before the interaction zone and therefore can be interpreted as a retrocausal and superdeterministic feature of the DS theory.  } \label{Fig3}
\end{figure}
Furthermore, the presence of radiated and advanced  $u$-field components propagating in the remote future or past, preserves the stability of the soliton at the same time as it preserves energy conservation. The old paradox associated with empty waves carrying and dissipating the corpuscle energy is therefore resolved in our time-symmetric DS approach \cite{Drezet2023}.\\   
\indent The situation is actually very general. Consider, as shown in Figure \ref{Fig3} a dBB particle scattered by an external classical potential located in a finite space-time region. Lets call $\mathcal{C}$ such a dBB trajectory crossing the interaction region. First, we note that the classical field can be impacted by actions coming from his past and we know from standard relativistic causality that such actions must be included in the past light cone  (here after denoted $\Delta$) having its apex on the interaction zone.  For example, if the external classical field can be switched or modified  we can  always imagine an external parameter  $\underline{\bold{a}}$ characterizing the mechanical or electromagnetic devices or settings associated  with the external field and that must be located in the relativistic causal past, i.e, in the backward light cone $\Delta$. Note that since the interaction region has a finite space-time extension we should rigorously consider several past light cones with apexes in the interaction zone: The next argument doest really needs that. Moreover, we can always find a configuration in which dBB positions $z(\lambda_{in})$ belonging to the trajectory  $\mathcal{C}$ and located before the interaction zone are causally independent from the  external field  and thus from   $\underline{\bold{a}}$ (that was clearly  the case in the example  of  Figure \ref{Fig2}). This will naturally occur when the wave packet associated with the incident wave function $\Psi(z(\lambda_{in}))$ is not overlapping or physically interacting with the devices characterized by the parameter  $\underline{\bold{a}}$  before the  interaction zone (ultimately if the parameter $\underline{\bold{a}}$ characterizes a light pulse coming from the past along the  light cone $\Delta$ there is no possibility--even in principle--to imagine an interaction between the hypothetically strongly localized wave function $\Psi(z(\lambda_{in}))$ and the mechanical or electromagnetic device $\underline{\bold{a}}$).\\
\indent Now, if we consider the point $z(\lambda_{in})$ we can calculate with Eq.~\ref{debroglieFFsoluSym} the $u-$field emitted into the past which is a function of the point $x$ and the position $z(\lambda_{in})$: $u_{adv}(x)=A(x,z(\lambda_{in}))$. Note that $x$ can not be arbitrary since information is constrained to propagate along the past light cone where $(x-z(\lambda_{in}))^2=0$ and $t:=x^0\leq z^0(\lambda_{in})$. Similarly, we can compute the $u-$ field radiated in the far future:   $u_{ret}(x)=R(x,z(\lambda_{in}))$ (see Figure \ref{Fig3}) with similar constraints along the future light cone.   These fields are causally independent from $\underline{\bold{a}}$ as it should be. However, the situation drastically changes after the interaction of the dBB particle with the external field.  The position $z(\lambda_{out})$ belonging to the dBB trajectory $\mathcal{C}$ after the interaction is indeed a function of the initial position and external field: 
\begin{eqnarray}
z(\lambda_{out}):=F_\Psi(\lambda_{out},z(\lambda_{in}),\underline{\bold{a}}).\label{inter}
\end{eqnarray} 
Obviously, the retarded and advanced fields emitted by the singularity in the far future or past will be functions of these parameters. Most importantly for us the advanced field emitted into the past reads 
\begin{eqnarray}
u_{adv}(x)=A(x,z(\lambda_{out}))=A_\Psi'(x,z(\lambda_{in}),\underline{\bold{a}})\label{alten}
\end{eqnarray} where we used Eq.~\ref{inter} and the  constraints $(x-z(\lambda_{out}))^2=0$, $t:=x^0\leq z^0(\lambda_{out})$ to remove the dependency over $\lambda_{out}$. Crucially, $u_{adv}(x)$ is now  depending on $\underline{\bold{a}}$. Therefore, along a space-like hyperplane $\Sigma(\lambda_{in})$ containing $z(\lambda_{in})$  (see Figure \ref{Fig3}) the field $u(x\in\Sigma(\lambda_{in}))$ which is located outside the limit provided by the intersection between $\Delta$ and $\Sigma(\lambda_{in})$  will depend on parameters such as $\underline{\bold{a}}$ even though  $z(\lambda_{in})$ is necessarily independent from these variables. In other words the description of the field is superdeterministic and retrocausal! It is interesting to add that  the total field $u(x)$ along  $\Sigma(\lambda_{in})$ will in general also contain a retarded contribution coming from points of the trajectory  $\mathcal{C}$ located much earlier $z(\lambda_{in})$ (not shown in Figure \ref{Fig3}). The $u-$field sum of retarded and advanced components is thus quite a complicated mathematical object which is strongly depending on the whole particle history. \\ 
\indent At that stage it is useful to give a very general discussion about causality in our DS approach. Here the $u-$field  at point $x$ is solution of the non linear equation Eq. \ref{1b} which equivalently reads 
\begin{eqnarray}
\Box u(x)=J(x):=\frac{3l_0^2}{(\frac{g_0}{4\pi})^4}(u(x)u^\ast(x))^2u(x)+\hat{\mathcal{O}}_x u(x)\label{1nnne}
\end{eqnarray} with the linear operator $\hat{\mathcal{O}}_x:=e^2A(x)^2-ie\partial_y A(x)-2ieA(x)\partial_x$. Using the Green theorem the formal solution reads 
\begin{eqnarray}
u(x)=\int_V d^4y K^{(0)}(x,y)J(y)+\oint_{\partial V}\varepsilon_y d^3S_yn_y\cdot[u(y)\partial_yK^{(0)}(x,y)-K^{(0)}(x,y)\partial_y u(y)]
\end{eqnarray} where $K^{(0)}(x,y)=K^{(0)}(y,x)$ is the Green propagator in vacuum [$d^3S_y$ is a three dimensional scalar elementary volume belonging to the boundary $\partial V$ at point $y$ and  $n_y$ is the outwardly oriented unit vector at point $y$  such that $\varepsilon_y=\textrm{sign}(n_y^2)\pm 1$]. If we consider the retarded Green function $K^{(0)}_{\textrm{ret}}(x,y)=\frac{\delta[(x-y)^2)]}{2\pi}\theta(x^0-y^0)$ the integral in Eq.~\ref{1nnne} can be pushed to the infinity and the surface integral along the boundary $\partial V$ includes only a contribution from a space-like hyperplane $\Sigma_{in}$ located in the remote past (i.e., at $t_y=y^0\rightarrow -\infty$). We thus have  
\begin{eqnarray}
u(x)=\int d^4y K^{(0)}_{\textrm{ret}}(x-y)J(y)+u_{in}(x)\label{integralecausale}
\end{eqnarray}  with the incident field $u_{in}(x)=-\int_{\Sigma_{in}}d^3\mathbf{y}[u(y)\partial_yK^{(0)}_{\textrm{ret}}(x,y)-K^{(0)}_{\textrm{ret}}(x,y)\partial_y u(y)]$. Using recursively Eq.~\ref{integralecausale} will allow us to express the total field $u(x)$ at a point $x$ as a functional 
\begin{eqnarray}
u(x)=\mathcal{F}(x;\{u_{in}(y),u^\ast_{in}(y), \partial u_{in}(y), \partial u^\ast_{in}(y)\}_{y\in \Delta_x})\label{function1}
\end{eqnarray} which depends on the incident fields $u_{in}(y),u^\ast_{in}(y), \partial u_{in}(y), \partial u^\ast_{in}(y)$ defined in the whole past light cone hypervolume $\Delta_x\in \mathbb{R}^4$ with apex at point $x$ (see Figure \ref{Fig4} (a)). Moreover, using the definition of $u_{in}$ we can restrict this definition to points $y$ located in the region of $\Delta_x$ located between $x$ and the past hypersurface $\Sigma_{in}$. We can also rewrite Eq.~\ref{function1} as  
\begin{eqnarray}
u(x)=\mathcal{G}(x;\{u(y),u^\ast(y), \partial u(y), \partial u^\ast\}_{y\in \Delta_x\cap\Sigma_{in}})\label{function2}
\end{eqnarray} 
which is a new functional depending only of the fields and derivatives along the part of the hyperplane $\Sigma_{in}$ included in the past light cone $\Delta_x$. 
\begin{figure}[h]
\includegraphics[width=8cm]{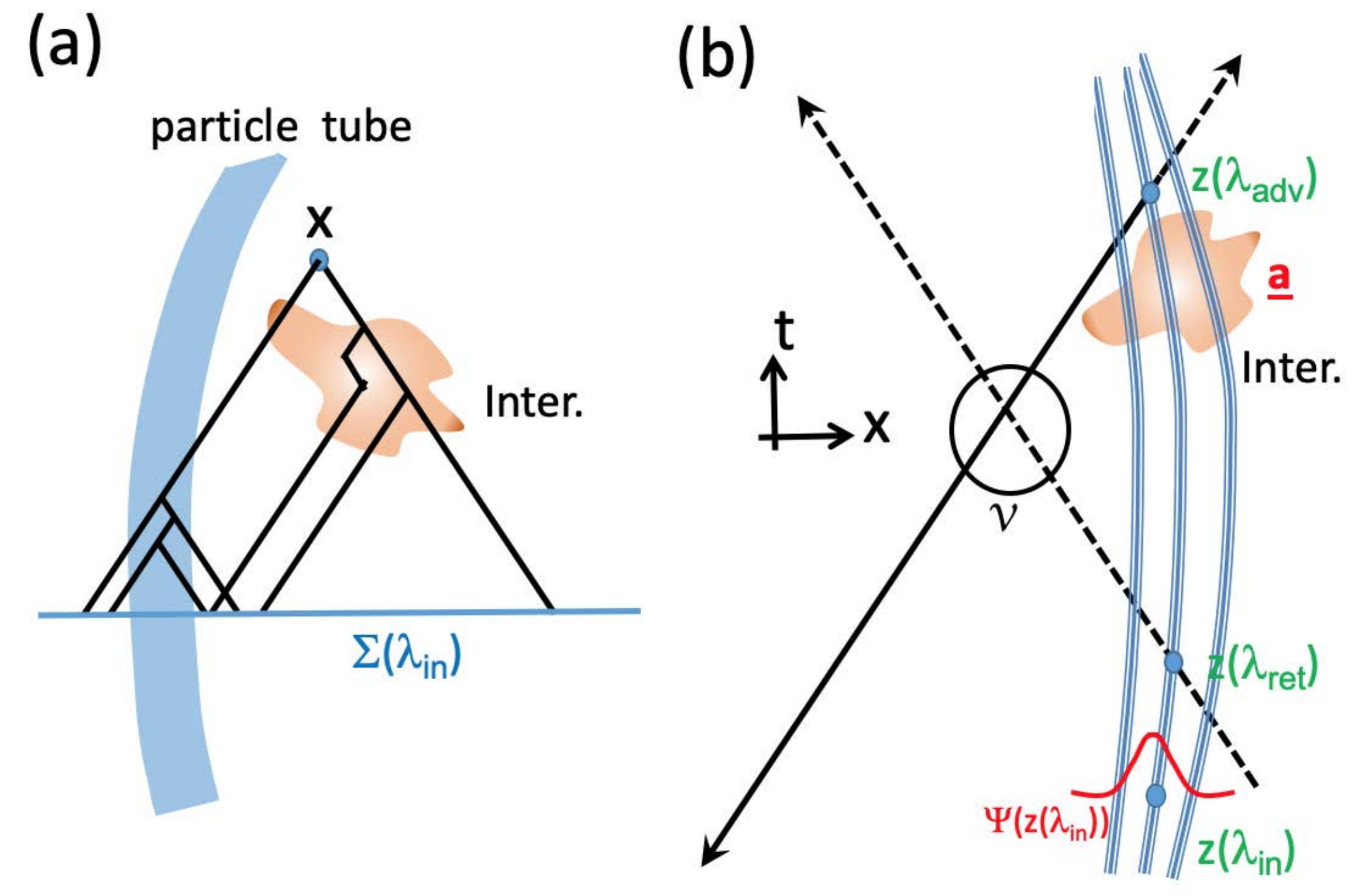} 
\caption{(a) A typical Cauchy problem where the $u-$field at point $x$ is expressed as functional of field incident field and scattered retarded field emitted in the backward moght cone $\Delta_x$. Here we show some typical scattering rays in presence of a particle tube (associated with a region of strong nonlinearity), and an interacting field. (a) Defining the probability $P(\{u(x)\}_{x\in\mathcal{V}})$ in the four dimensional  volume $\mathcal{V}$ requires the knowledge of retarded and advanced $u-$fields emitted by the soliton during its dBB motion. Moreover the dBB statistical distribution  associated with the initial wave function  $\Psi(z(\lambda_{in}))$ will weight this otherwise completely deterministic property defined in volume $\mathcal{V}$. Advanced and retarded contributions to the $u-$field lead to a mere violation of local causality involving only past light-cones.}\label{Fig4}
\end{figure}
This naturally defines the Cauchy problem where the knowledge of initial conditions of the fields and derivatives  on 
$\Delta_x\cap\Sigma_{in}$ is necessary and sufficient to compute (in principle algorithmically) the field $u(x)$ at the apex of $\Delta_x$. However, in the present theory the evolution equation Eq.~\ref{1nnne} is strongly nonlinear and nonlinear equations are difficult to solve.  It would be difficult to guess by inserting some input fields $u_{in}$ by hand what would be the final solution and in particular if this would lead to a stable soliton. What we showed is that the theory admits self-consistent solitonic solutions having a time symmetric structure $u_{\textrm{sym}}(x)=\frac{u_{\textrm{ret}}(x)+ u_{\textrm{adv}}(x)}{2}$. As we already explained in Section\ref{sub32} we can always equal this solution to the natural Cauchy  solution Eq.~\ref{integralecausale}  written as $u_{\textrm{sym}}(x)=u_{\textrm{ret}}(x)+u_{\textrm{in}}(x)$ if we put $u_{\textrm{in}}(x)=\frac{u_{\textrm{adv}}(x)-u_{\textrm{ret}}(x)}{2}$. But now because of the presence of advanced fields in its definition $u_{\textrm{in}}(y)$ computed in  $\Delta_x$  will depend on the future field at point $x$. Generally speaking systems of equations  involving retrocausal links can lead to mathematical inconsistencies  due to causal loops. Here however, we found a family of self consistent solitons driven by dBB trajectories $z_i(\lambda)$ guided by a $\Psi-$field solution of the linear Klein-Gordon or Schr\"odinger equation. The self consistency means that if we used the $u-$field computed from our solitons to define the input fields variables needed in Eqs.~\ref{function1},\ref{function2} we could in principle check that the $u-$field at position $x$ could be precisely recomputed to give the input field that is required... we get a kind of algorithmic loop! but a self-consistent one unlike the infamous `grand-father paradox' where a grandson acting backward in time could kill his own grand-father long time before his own birth. \\
\indent At a fundamental and more philosophical level this leads to interesting questions if we try to identify the whole Universe to a kind of computer calculating algorithmically the $u-$field at any position $x$ as a function (or functional) of its causal past (i.e., the past included in $\Delta_x$). We see that the Cauchy approach going traditionally from past to future would not be a good  or efficient one without knowing already in advance the solitonic solutions of our nonlinear equations. At a cosmological level, i.e., with a Big-Bang this would even require a fine tuning or conspiratorial scenario. Moreover, in a block Universe picture taking seriously the symmetry of nature between space and time dimensions  the fact to use self-consistent time-symmetric fields is not ridiculous. Clearly, however it leads to interesting questions concerning causality, superdeterminism, and free-will.\\
\indent One of this question concerns the concept of probability in our DS theory. Consider for example the case sketched in Figure \ref{Fig4} (b) where a particle interacts with an external field characterized as before by a parameter  $\underline{\bold{a}}$.   Knowing the initial dBB position distribution $dP_\Psi(z(\lambda_{in}))$, given by the wavefunction $\Psi(z(\lambda_{in}))$ and Born's rule, will allow us to define the probablity for the field $u(x)$ to have some specified  values in the four dimensional volume  $\mathcal{V}$. Writing $P(\{u(x)\}_{x\in\mathcal{V}})$ this probability we have 
\begin{eqnarray}
P(\{u(x)\}_{x\in\mathcal{V}})=\int dP_\Psi(z(\lambda_{in}))\delta(\{u(x)-u_{\textrm{sym}}(x)\}_{x\in\mathcal{V}})
\end{eqnarray} where the integration is done over the dBB particle  distribution $dP_\Psi(z(\lambda_{in}))$, and where the Dirac distribution is a a functional (required because the theory is deterministic) and where $u_{\textrm{sym}}(x)=\frac{u_{\textrm{ret}}(x)+u_{\textrm{adv}}(x)}{2}$ depends as before on the history of the dBB particle. In particular the $u-$field in region $\mathcal{V}$ can clearly depend on the position of the particle in the future light cone  with apex at point $x$ due to the presence of advanced wave components $u_{\textrm{adv}}(x)$ (see Figure \ref{Fig4} (b)). The probability $P(\{u(x)\}_{x\in\mathcal{V}})$ therefore violates the local causality requirement of Bell. Interestingly this also violates the idea of the usual dBB theory that a probability  can not depend on future events (Lucien Hardy and Squires called this  the principle of outcome independence from  later measurements: POILM \cite{Hardy,DrezetFP2019}).
Moreover, POILM was adapted to the dBB pilot-wave theory and concerned particle observables associated with the particle presence at point $z$. Here we are considering  the $u-$field at points $x\neq z$ in the context of the DS theory.  Furthermore, it is not required to consider $u(x)$ as a physical observable in general in the sense that the detection of a particle presupposes that the core of the soliton with a highly nonlinear $u-$field is crossing the region $\mathcal{V}$ which is not the case in the example discussed here where only a weak far-field is supposed to reach region $\mathcal{V}$. $P(\{u(x)\}_{x\in\mathcal{V}})$ is better interpreted as a probability concerning  (hidden) beables or ontic states of the quantum system described by the DS theory.  An application concerns the case where instead of the volume   $\mathcal{V}$ we consider a part $\delta \Sigma_{in }$  of the Cauchy hypersurface $\Delta_x\cap\Sigma_{in}$ (i.e, $\delta \Sigma_{in }\subseteq\Delta_x\cap\Sigma_{in}$) associated with the causal past of the particle. We have thus
\begin{eqnarray}
P(\{u(x)\}_{x\in\delta \Sigma_{in }})=\int dP_\Psi(z(\lambda_{in}))\delta(\{u(x)-u_{\textrm{sym}}(x)\}_{x\in\delta \Sigma_{in }})
\end{eqnarray} which shows that the incident field needed to apply the Cauchy problem in Eq.~\ref{function2} is itself associated with a probability distribution deduced from the dBB distribution $dP_\Psi(z(\lambda_{in}))$. Ultimately, using Eq.~\ref{function2} the $u-$field near the particle singularity: $u(x\simeq z(\lambda)):=u_{\textrm{sym}}(x\simeq z(\lambda))$ is itself associated with a probability as it should be in order for the DS and dBB theory to be self-consistent.\\
\indent  A central point in our analysis is that we obtained two different alternative descriptions of the $u-$field:  (A) in the one side, we have the usual Cauchy description involving past  information over the hyper-surface $\Delta_x\cap\Sigma_{in}$ (see Eqs.~\ref{function1},\ref{function2}). This description would be very difficult to use in practice for a nonlinear field. (B)  On the other side, we have the time-symmetric description used in the present work leading,  e.g., to  Eq.~\ref{debroglieFFsolu} or Eq.~\ref{soluN} in the far-field; the near-field beeing described by Eqs.~\ref{Lanesolutionscaled},\ref{Lanesolutionscaledmany}. This new description relies on the knowledge of dBB particle paths  $z_j(\lambda)$ which can be expressed as functions of the initial dBB coordinates $z_j(\lambda_{in})$. By integration we have $z_j(\lambda)=F_{\Psi_N,j}(\lambda,\{z_k(\lambda_{in})\})$. Importantly, these dBB trajectories are not depending on future events as it was clearly analyzed by Hardy and Squires in \cite{Hardy} (see also \cite{DrezetFP2019}) with POILM. This is exactly what is happening in the example of Eq.~\ref{inter} which depends on the parameter $\underline{\bold{a}}$ only after the interaction of the particle with the external field. By inserting these expressions for  $z_j(\lambda)$ into $u_{\textrm{sym}}(x)$ given by Eq.~\ref{soluN} we thus obtain formulas like Eq.~\ref{alten} for the $u-$field which most generally would read:
\begin{eqnarray}
u_{\textrm{sym}}(x):=\frac{1}{2}R_{\Psi_N}(x,\{z_k(\lambda_{in})\})+\frac{1}{2}A_{\Psi_N}(x,\{z_k(\lambda_{in})\})\label{machin}
\end{eqnarray} where the retarded field $R_{\Psi_N}$ and the advanced field  $A_{\Psi_N}$ depend in general on the particle histories and interactions and constrained by the Hardy/Squires dBB causality principle POILM. In the end the time-symmetric field of  Eq.~\ref{machin} with the dBB input variables $\{z_k(\lambda_{in})\}$ is rigorously equivalent to the Cauchy description of Eq.~\ref{function2}. However, the time-symmetric description requires much less local hidden variables or beables for its description since ultimately it requires only the dBB coordinates $\{z_k(\lambda_{in})\}$  and not the full knowledge of the fields  $\{u(y),u^\ast(y), \partial u(y), \partial u^\ast\}_{y\in \Delta_x\cap\Sigma_{in}}$ over the Cauchy hypersurface.\\ 
\indent The previous analysis applies to the  last problem that we must discuss here namely Bell's theorem and nonlocality. Indeed, it is remarkable that our local DS  theory, as shown in Section~\ref{sec4}, allows for a description of $N$ solitons involving a $\Psi-$function $\Psi_N(\{z_j(\lambda)\}) $ associated with  $N$ entangled dBB particles. Indeed, $\Psi_N(\{z_j(\lambda)\}) $ beeing associated with the Klein-Gordon equation admits solutions having a strong nonlocal character in the sense that these solutions can be used to violate some Bell's inequalities. The dBB trajectories obtained with the guidance formula Eq.~\ref{PWIguidanceGenN} are thus strongly correlated and the particles,  characterized by the varying masses $\mathcal{M}_{\Psi_N,j}(\{z_j(\lambda)\})$ ascing as relativistic quantum potential, are submitted to nonlocal instantaneaous forces. Altogether, this violates the conditions of local-causality and or statistical independence  defined by Bell and reminded in  Section \ref{sec2}. To recap once more: The DS theory is fundamentally relativistically local (even though nonlinear) whereas the dBB pilot-wave theory is  nonlocal and requires a preferred foliation (as discussed for example in \cite{DrezetFP2019}) or synchronisation (as discussed in Section \ref{sec4}).  There is thus a clear tension between the dBB  pilot-wave theory  and the DS theory developed here. How can we solve this apparent contradiction?\\
\indent The central idea to solve this dilemma is to take seriously the time-symmetric $u-$field of our DS theory. Indeed, from Eq.~\ref{soluN}     we deduce that the solitons or singularities emit advanced waves that propagate backward in time and can in turn carry information from the futur to the past. This retrocausal link can be used to define the incident $u-$field  along a past Cauchy surface $\Sigma_{in}$. In turn we have thus a way to decipher the mysterious nonlocal link between particles using  time-symmetry to justify a form of superdeterminism.\\
\begin{figure}[h]
\includegraphics[width=10cm]{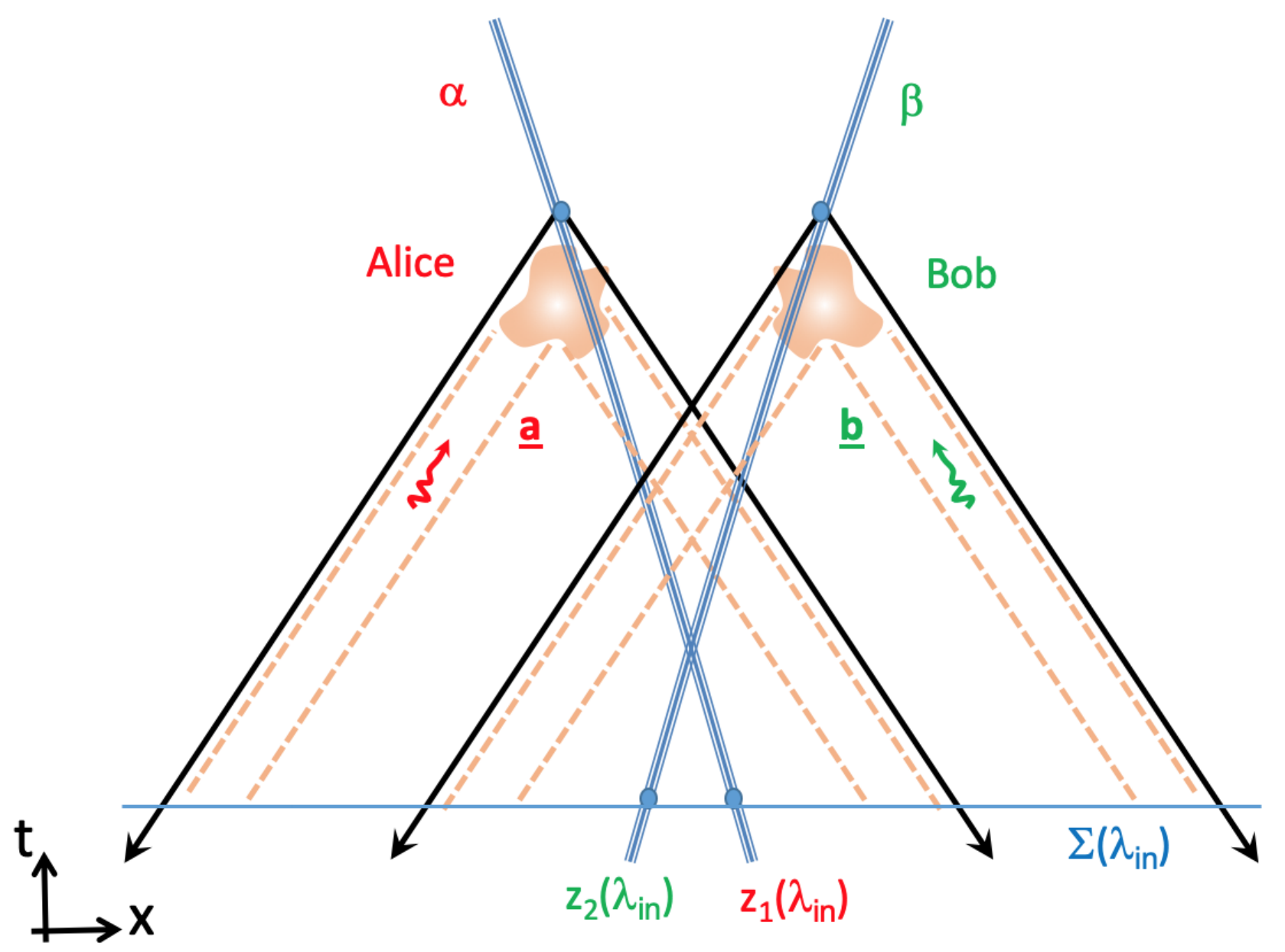} 
\caption{Bell's scenario involving two dBB particles associated with solitons.}\label{Fig5}
\end{figure}
\indent To be more precise consider for example, as  shown in Figure \ref{Fig5}, two entangled particles forming an EPR pair (systems of spinless Klein-Gordon particles having this property are discussed in \cite{DrezetFP2019}). We suppose that the two entangled particles are sent to observers Alice and Bob located in remote labs where fields act locally on the separated particles. The settings  $\underline{\bold{a}}$ (respectively  $\underline{\bold{b}}$) associated  with the external fields acting on particle  1  (respectively 2) are for example driven by random optical signals coming from remote stars or quasars as in \cite{Zeilinger2,Zeilinger3,Kaiser,Zeilinger4}. Alice and Bob operations on the aprticle 1 and 2 lead to measurements of dichotomic observables $\alpha$ and $\beta$ that can take values $\pm 1$. Assuming that there is no  other `observable and physical'  signals coming from the past light cones with apexes in the interaction regions (see  Figure \ref{Fig5}) the particles detected by Alice and Bob can not know in advance the settings and operations realized by Alice and Bob. Therefore, from the point of view of Bell or Einstein, the particles 1 and 2 could not be used to violate a Bell test. Still they do of course, as it has been experimentally checked many times \cite{Aspect,Zeilinger,Hanson},  and this means that other information is needed in the backward light cones to satisfy the principle of local causality of Bell and Einstein.  The only solution, if one wants to preserve the results of special relativity, is apriori to relax the condition \ref{c} of Section \ref{sec2}, i.e.,  $dP_{12}(\boldsymbol{\Lambda}|\mathbf{\underline{a}},\mathbf{\underline{b}})=dP_{12}(\boldsymbol{\Lambda})$ for the beables $\boldsymbol{\Lambda}$  and, in other words, to abandon statistical independence. But, would complain a Bohmian, this contradicts the assumptions of the dBB pilot-wave theory where Bell beables are  the particle initial coordinates $\boldsymbol{\Lambda}\equiv [z_1(\lambda_{in}),z_2(\lambda_{in})]$. In our theory we assume the results dBB  theory and we cannot apriori `it seems' save local causality.  This is the case since the dBB particles crossing the interaction zones of Alice and Bob are coupled by a nonlocal link defined in the phase $S_2(z_1(\lambda),z_2(\lambda))$ of the wave function $\Psi_2(z_1(\lambda),z_2(\lambda))$. It is this nonlocal link that produces the instantaneous action at a distance between the particles which in turn violates at least one of the two local-causality conditions \ref{a} and \ref{b} of Section \ref{sec2}. In other words,  following the dBB pilot-wave theory we will have in the interaction zones and after:
  \begin{eqnarray}
z_1(\lambda_{out}):=F_{1,\Psi_{12}}(\lambda_{out},z_1(\lambda_{in}),z_2(\lambda_{in}),\underline{\bold{a}},\underline{\bold{b}})\nonumber\\
z_2(\lambda_{out}):=F_{2,\Psi_{12}}(\lambda_{out},z_1(\lambda_{in}),z_2(\lambda_{in}),\underline{\bold{a}},\underline{\bold{b}}).
\label{inter12}
\end{eqnarray} the dBB trajectories are thus nonlocally depending on both settings  $\mathbf{\underline{a}},\mathbf{\underline{b}}$.\\ 
\indent Moreover, in  our DS theory the $u-$field of the two singularities reads:
\begin{eqnarray}
u_{\textrm{sym}}(x)=g_0\int d\lambda\sqrt{|\dot{z}_1(\lambda)\dot{z}_1(\lambda)|} K_{\textrm{sym}}(x,z_1(\lambda))
\frac{e^{iS_{12}(z_1(\lambda),z_2(\lambda))}}{\sqrt{\alpha_1(z_1(\lambda),z_2(\lambda))}}\nonumber\\
+g_0\int d\lambda\sqrt{|\dot{z}_2(\lambda)\dot{z}_2(\lambda)|} K_{\textrm{sym}}(x,z_2(\lambda))
\frac{e^{iS_{12}(z_1(\lambda),z_2(\lambda))}}{\sqrt{\alpha_2(z_1(\lambda),z_2(\lambda))}}.\label{soluEPR}
\end{eqnarray}
In particular, this time-symmetric field can be used to compute the advanced waves propagating along the backward light cones with apexes in the interaction zones of Alice and Bob. These waves carry precisely in the overlap of the two light cones the physical information on both settings $\mathbf{\underline{a}},\mathbf{\underline{b}}$  that was supposed not to exist in the standard dBB theory. Therefore, along a Cauchy surface $\Sigma_{in}$ and more precisely in the overlap of such a surface  with the region included in the past light cones we have information about both settings $\mathbf{\underline{a}},\mathbf{\underline{b}}$, i.e, about information occuring in the future of the solitons with respect to  $\Sigma_{in}$.\\
\indent The situation is even more explicit if we consider the outcomes $\alpha, \beta$ which reads in the dBB theory:
\begin{eqnarray}
\alpha=A_{\Psi_{12}}(z_1(\lambda_{in}),z_2(\lambda_{in}),\underline{\bold{a}},\underline{\bold{b}}):=A_{\Psi_{12}}(\boldsymbol{\Lambda},\underline{\bold{a}},\underline{\bold{b}})\nonumber\\
\beta=B_{\Psi_{12}}(z_1(\lambda_{in}),z_2(\lambda_{in}),\underline{\bold{a}},\underline{\bold{b}}):=B_{\Psi_{12}}(\boldsymbol{\Lambda},\underline{\bold{a}},\underline{\bold{b}})\label{model1}
\end{eqnarray}  and depend nonlocally on the settings $\underline{\bold{a}},\underline{\bold{b}}$ and the local beables $\boldsymbol{\Lambda}\equiv [z_1(\lambda_{in}),z_2(\lambda_{in})]$. These are the quantities involved in the Alice and Bob join measurements  leading to the violation of Bell inequalities. Once more, for a Bohmian the situation is clearly demonstrating the necessary non local link based on an instantaneous action at a distance. However, in the present DS theory we have alternatively after using Eq.~\ref{function2}:   
\begin{eqnarray}
\alpha=\mathcal{A}\left(\{u(y),u^\ast(y), \partial u(y), \partial u^\ast\}_{y\in (\Delta_1\cap\Sigma_{in})\cup (\Delta_2\cap\Sigma_{in})}\right)\nonumber\\
\beta=\mathcal{B}\left(\{u(y),u^\ast(y), \partial u(y), \partial u^\ast\}_{y\in (\Delta_1\cap\Sigma_{in})\cup (\Delta_2\cap\Sigma_{in})}\right)\label{model2}
\end{eqnarray} which defines Cauchy functional requiring the knowledge of field variables $u(y),u^\ast(y), \partial u(y), \partial u^\ast$ over the hypersurface $(\Delta_1\cap\Sigma_{in})\cup (\Delta_2\cap\Sigma_{in})$ (associated with the intersection of the light cone regions $\Delta_{1,2}$ with apexes on detector settings of particles 1 and 2) with the whole Cauchy hyperplane $\Sigma_{in}$. This causal region is the only one which is physically relevant for evaluating observables $\alpha$ and $\beta$. In this description there is no nonlocal link: Fundamentally everything is local and obeys to to Eq.~\ref{1b}. Moreover, the advanced wave contribution to the  time-symmetric solution Eq.~\ref{soluEPR} allows us to compute retrocausally $u(y),u^\ast(y), \partial u(y), \partial u^\ast$ over the hypersurface $(\Delta_1\cap\Sigma_{in})\cup (\Delta_2\cap\Sigma_{in})$. Since the advanced field depends through Eq.~\ref{inter12} on the input variables $\boldsymbol{\Lambda}\equiv [z_1(\lambda_{in}),z_2(\lambda_{in})]$ and the parameters $\mathbf{\underline{a}},\mathbf{\underline{b}}$ we have a superdeterministic theory.\\
\indent  In other words,  Eqs.~\ref{model1} and \ref{model2} are alternative but equivalent descriptions of the observables $\alpha,\beta$. In the DS theory based on the $u-$field the nonlocality of the dBB pilot-wave theory is not fundamental. Rather, it  constitutes and effective description which ignores the time-symmetric motion of the $u-$wave and focuses only on the entangled motion of the solitons cores. If we ignore the $u-$waves and only watch the dBB trajectories we miss the causal information associated with advanced and retarded components of the $u-$field that are converging on the particles. If we alternatively watch the motion of the $u-$field  and apply a Cauchy-like perspective where the time is flowing from past to future  we will see a superdeterministic theory where the incident field  over $(\Delta_1\cap\Sigma_{in})\cup (\Delta_2\cap\Sigma_{in})$ depends on the future states through the parameters $\underline{\bold{a}},\underline{\bold{b}}$.  Somebody having acces to this initial $u-$field would thus conclude to a conspiratorial scenario where the field is tuned exactly in the precise way to reproduce the dBB nonlocal motions and the predictions of quantum mechanics, e.g.,  the Bell inequalities violations.\\
\indent In the recent debates concerning quantum foudations and the Bell theorem superdeterminism has been often neglected or considered as a `absurd' solution that would bring into question the whole science methodology assuming statistical independence (for interesting exceptions and attempts see however \cite{Hooft,Palmer,Ciepielewski}). This could apriori question the values of `free choice' done by Agents like Alice and Bob.  
However, the methodology assuming statistical independence  was established for macroscopic phenomena. As we showed the DS theory developed initially by de Broglie and pushed here to its logical conclusion requires time-symmetric fields and this ultimately justifies a form of superdeterminism to recover the guidance formula of the dBB theory. This superdeterminism is however not like the one of a magician on the stage using  tricks to  mistake us and create an illusion. The fine-tuning needed in our theory is not so contingent as for the magician since it  is needed to justify the existence and stability of the time-symmetric solitons. Indeed, the nonlinearity of the DS theory forces us to assume a time-symmetric $u-$field and in turn this time-symmetry is essential to explain why the superdeterminism along the Cauchy hyperplane $\Sigma_{in}$ located in the past is caused by the properties of the solitons in the future. Furthermore, there is no question concerning free-choice or free-will  here and this for at least two reasons. First, like  classical and Bohmian mechanics, the  DS theory is fully deterministic  and the concept of free-will is necessarily an illusion even if very useful (we will not here enter the philosophical debates betweeen incompabilists and compabilists concerning free-will and determinism). Second, we should not forget that our theory agrees with the dBB pilot-wave theory concerning the predictions and statistics of quantum events. In an experiment like the one proposed before with an EPR pair we naturally suppose that the $\Psi-$wave functions associated with the two observers, i.e.,  $\Psi_A$ and $\Psi_B$, factorize from the EPR state $\Psi_{12}$ before the interactions and measurements in the remote labs.  this fact implies that the dBB trajectories associated with the observers and the pair of particles are not correlated and the sub-systems are statistically independent.  The DS theory will in general involves advanced fields that in the usual time direction (i.e., from past to future) are converging from the remote past with information about settings $\mathbf{\underline{a}},\mathbf{\underline{b}}$. This superdeterministic information will converge on the particles and observers only during and after the interactions. Before these convergence the observers were not influenced by any superdeterminstic information and their actions were not less no more free than in Newtonian mechanics. In the end everything is consistent with the dBB theory interpreting the observer journeys  through nonlocal links manifesting during the interactions.  \\
\indent To conclude this long story, it is probably useful to remind once again that nonlinear field equations admitting solitons are extremelly difficult to solve and probably these studies are still in their infancy. The general methodological message of the present study is therefore that we should perhaps not be so astonished that  new features require to relinquish prejudices about causality and time-symmetry. Progresses in science are often made by abandoning prejudices. This is specially true since our theory is fundamentally relativistic and requires to take seriously the symmetry existing between space and time. This was clearly the path followed by de Broglie in 1925 when he attempted to explain quantum mechanics using the DS theory. Here, by taking this path seriously we developed a complete DS theory able to reproduce a large set of quantum features associated with the Klein-Gordon relativistic equation.  The DS theory in turn explains wave-particle dualism and recovers the predictions of the dBB pilot-wave theory for relativistic particles in external electromagnetic fields.  The DS theory also explains the nonlocality and the action at a distance of the dBB pilot-wave theory as an effective description. Fundamentally there is however no such `spooky' action at a distance entering into conflict with special relativity. The DS theory, thanks to time-symmetry is local but superdeterministic.       
\appendix
\section[\appendixname~\thesection]{The tachyonic regime for solitons}
\label{appendix}
\indent As it is well known the dBB theory for the Klein-Gordon equation leads to some paradoxical features already discussed  in 1927 \cite{Valentini,debroglie1930} in the regime where  $\mathcal{M}^2_\Psi(x)=(\partial S(x)+eA(x))^2\leq 0$. In this regime the mass becomes imaginary and this implies a faster than light, i.e., tachyonic, motion   for the particle guided by the $\Psi-$wave.  However, contrarily to some old claims (see for example \cite{Kyprianidis}) the dBB pilot-wave theory can be developed self-consistently even in the regime where   $\mathcal{M}^2_\Psi(x)\leq 0$. For this purpose it is enough to generalize Eq.~\ref{PWIguidance} as 
\begin{eqnarray}
\frac{dz(\lambda)}{d\lambda}=-(\partial S(z(\lambda))+eA(z(\lambda))\sqrt{\left(\frac{\frac{dz_\mu(\lambda)}{d\lambda}\frac{dz^\mu(\lambda)}{d\lambda}}{\mathcal{M}^2_\Psi(z(\lambda))}\right)}\label{PWIguidanceGen}
\end{eqnarray} where $\lambda\in \mathcal{R}$ is a parameter evolving along the particle trajectory. If  $\mathcal{M}^2_\Psi(z)\geq 0$ this must be used with $\frac{dz_\mu(\lambda)}{d\lambda}\frac{dz^\mu(\lambda)}{d\lambda}\geq 0$ corresponding to 	a time-like motion. However if $\mathcal{M}^2_\Psi(z)\leq 0$ we must take $\frac{dz_\mu(\lambda)}{d\lambda}\frac{dz^\mu(\lambda)}{d\lambda}\leq 0$ associated with a space-like, i.e., tachyonic regime  for the particle.\\
\indent In the case of a space-like motion with $d\tau^2=dz_\mu(\lambda)dz^\mu(\lambda)<0$ we can use the parameter $\sqrt{d\tau^2}=id\theta$, i.e., $\sqrt{-d\tau^2}=d\theta\in \mathbb{R}$ along the trajectory. Similarly since $\mathcal{M}^2_\Psi(z)\leq 0$ we can introduce $\sqrt{\mathcal{M}^2_\Psi(z)}=i\Omega_\Psi(z)$, i.e.,  $\sqrt{-\mathcal{M}^2_\Psi(z)}=\Omega_\Psi(z)\in \mathbb{R}$. These definitions correspond to a continuation in the complex plane of the subluminal formula defined in Eq.~\ref{PWIguidance}. We have thus 
\begin{eqnarray}
\frac{d z(\theta)}{id\theta}:=-iv_\Psi(z(\theta))=-\frac{\partial S(z(\theta))+eA(z(\theta))}{i\Omega_\Psi(z(\theta))}\label{PWIguidanceCo}
\end{eqnarray} i.e., $\frac{d z(\theta)}{d\theta}=-\frac{\partial S(z(\theta))+eA(z(\theta))}{\Omega_\Psi(z(\theta))}$ in agreeement with Eq.~\ref{PWIguidanceGen}. It is important to note that along a dBB trajectory a subluminal segment (with $\mathcal{M}^2_\Psi(z)\geq 0$) is necessarily separated from a superluminal segment (with $\mathcal{M}^2_\Psi(z)\leq 0$) by a  critical point where the mass vanishes, i.e.,  $\mathcal{M}^2_\Psi(z)=0$. This explains why in the dBB  theory a particle can cross the light-cone.  In general tachyonic sectors are confined to finite regions of space-time and are associated  with evanescent $\Psi-$fields or occur in interference zones near fringes minima as already shown by de Broglie \cite{debroglie1930}. Moreover, this also explains why tachyonic motions are not problematic from the causal point of view since the tachyonic solutions are not observed in the the far-field, i.e., remotly from interference zones or strong scattering potentials.\\ 
\indent In the context of the DS theory developed here we need to define an hyperplane locally orthogonal to the trajectory. In the time-like subluminal regime this space-like hyperplane $\Sigma(\tau)$  is defined by the condition $\xi\cdot\frac{d z(\tau)}{d\tau}=0$ with $\xi=x-z(\tau)$  that corresponds to the local inertial rest frame following the particle motion  $z(\tau)$. In the tachyonic regime we can similarly define a local hyperplane $\Sigma(\theta)$  by the condition $\xi\cdot\frac{d z(\theta}{d\theta}=0$ with $\xi=x-z(\theta)$ but $\Sigma(\theta)$ is not space-like.  In order to physically interpret this hyperplane we consider a laboratory frame where the particle is at time $t$ moving with a velocity $v>1$ along the $+x$ spacial direction. We thus define a generalized Lorentz transformation $x'=-i\frac{(x-vt)}{\sqrt{v^2-1}}:=-ix''$, $t'=-i\frac{(t-vx)}{\sqrt{v^2-1}}:=-it''$ (with $x'',t''\in \mathbb{R}$), $y'=y:=y''$, $z'=z:=z''$. The hyperplane $\Sigma(\theta)$ corresponds to $t''=$Const. Equivalently, we can define a Lorentz frame moving with the subluminal velocity $w=1/v<1$ along the $+x$ spacial direction and introduce coordinates $x'''=\frac{(x-wt)}{\sqrt{1-w^2}}:=-t''$, $t'''=\frac{(t-wx)}{\sqrt{1-w^2}}:=-x''$, $y'''=y$, $z'''=z$. Interestingly, we see that comparing the coordinates the spacial coordinate  $x''$ corresponds to the time  $-t'''$ and conversely the time $t''$ corresponds to the spacial coordinates $x'''$.  These  equivalent sets of new coordinates characterize the hyperplane $\Sigma(\theta)$ by  $t''=-x'''$=Const. Hence, the DS approach developed in Section \ref{sub31} is easily extended to the tachyonic case. First, we can similarly to Eq.~\ref{phaseharmony}, define in the hyperplane $\Sigma(\theta)$ the phase harmony condition  
\begin{eqnarray}
\varphi(x) \simeq S(z(\theta)) -eA(z(\theta))\xi+B(z(\theta))\frac{\xi^2}{2}+O(\xi^3).\label{phaseharmonyGen}
\end{eqnarray}  From this we recover the guidance formula Eq.~\ref{DSguidance}
\begin{eqnarray}
\frac{d z(\theta)}{d\theta}=v_\Psi(z(\theta))=v_u(z(\theta))\label{DSguidanceG}
\end{eqnarray}
with $v_u(x)=-\frac{\partial \varphi(x)+eA(x)}{\sqrt{-(\partial \varphi(x)+eA(x))^2}}$ and from the continuity equation we generalize Eq.~\ref{newdfluid} as 
\begin{eqnarray}
-\partial v_u(z(\theta)):=-\frac{d}{d\theta}\ln{[\delta^3\sigma_0(z(\theta))]}
=\frac{d}{d\theta}\ln{[f^2(z(\theta))\Omega_\Psi(z(\theta))]}=\frac{3B(z(\theta))}{\Omega_\Psi(z(\theta))}.\label{newdfluidGe}
\end{eqnarray} The soliton profile is determined in the hyperplane $\Sigma(\theta)$ by the equation:  
\begin{eqnarray}
\Omega^2_\Psi(z(\tau))f(x)+(\frac{\partial^2}{\partial {x''} ^2}-\frac{\partial^2}{\partial y ^2}-\frac{\partial^2}{\partial z ^2})f(x)\simeq -\frac{3l_0^2}{(\frac{g_0}{4\pi})^4}f^5(x))
\label{ODEGE}\end{eqnarray}  
which admits in the near-field the solution
\begin{eqnarray}
f(x):=G_\alpha(x,z,x'')=\frac{\sqrt{\alpha}g_0}{4\pi}\frac{1}{\sqrt{\alpha^2(y^2+z^2-(x''+i\Delta)^2)+l_0^2}}.\label{LanesolutionscaledG}
\end{eqnarray} The parameter $\Delta \in\mathbb{R}$ is introduced in order to remove the singularity that would otherwise appear on the hyperboloids $y^2+z^2-{x''}^2=-(l_0/\alpha)^2$. We stress that this mathematical expression is reminiscent of X-type superluminal waves discussed in optics \cite{Saari}.     Moreover, for practical reasons it is possible to neglect the role of $\Delta$. Indeed, if we consider a continuous dBB trajectory the subliminal and superluminal regions are separated by    critical points where $\mathcal{M}^2_\Psi(z)=0$.  But as we showed in Section \ref{sub31} a soliton starting from a subluminal path coming from past infinity (i.e., $\tau=-\infty$) is characterized by a coefficient $\alpha(\tau)=\alpha(-\infty)\sqrt{[\mathcal{M}_\Psi(z(\tau))/\mathcal{M}_\Psi(z(-\infty))]}$ which is vanishing at the critical point. In other words, from Eqs.~\ref{Lanesolutionscaled}, and \ref{const} we deduce $l(z(\tau))\rightarrow +\infty$, $f(z(\tau))\rightarrow 0$ when we approach the critical point from the subluminal side.   On the the other side, i.e., in the tachyonic regime, we must by continuity als0 impose that $\alpha\rightarrow 0$ in Eq.\ref{LanesolutionscaledG} near the critical point. Therefore, the singular hyperboloids $y^2+z^2-{x''}^2=-(l_0/\alpha)^2$ would be located at infinity and are not physical. For this reason,  we can fairly assume $\Delta=0$ as a sufficient approximation for describing the soliton near-field.\\ 
\indent Ultimately, using the fluid conservation law Eq.~\ref{2d} and Eq.~\ref{newdfluidGe} we  easily obtain the general evolution of  $\alpha(z(\lambda))$ and $B(z(\lambda))$ along the trajectory involving tachyonic and subluminal segments:
\begin{eqnarray}
\alpha(z(\lambda))=\alpha(z(-\infty))\sqrt{\left(\frac{|\mathcal{M}_\Psi(z(\lambda))|}{|\mathcal{M}_\Psi(z(-\infty))|}\right)}=\sqrt{\left(\frac{|\mathcal{M}_\Psi(z(\lambda))|}{\omega_0}\right)}\\
B(z(\lambda))=\frac{1}{2\sqrt{\left(\frac{dz_\mu(\lambda)}{d\lambda}\frac{dz^\mu(\lambda)}{d\lambda}\right)}}\frac{d \sqrt{\mathcal{M}^2_\Psi(z(\lambda))}}{d\lambda}\label{inteGSu}
\end{eqnarray}  
  where we assumed the asymptotic values $\alpha(z(-\infty))=1$ and $\mathcal{M}_\Psi(z(-\infty))=\omega_0$. In particular, in the subliminal regime we recover Eq.~\ref{inteG} and in the superluminal regime we have  $B(z(\theta))=\frac{1}{2}\frac{d}{d\theta}\Omega_\Psi(z(\theta)) $.



\begin{thebibliography}{}
\bibitem{debroglie1923a}
De Broglie, L. Ondes et quanta.  \textit{C.~R. Acad.~Sci. (Paris)} \textbf{1923}, \textit{177}, 507--510.
\bibitem{debroglie1923b}
De Broglie, L. Quanta de lumi\`{e}re, diffraction et interf\'erences.  \textit{C.~R. Acad.~Sci. (Paris)} \textbf{1923}, \textit{177}, 548--560.
\bibitem{debroglie1923c}
De Broglie, L. Les quanta, la th\'eorie cin\'etique des gaz et le principe de Fermat.  \textit{C.~R. Acad.~Sci. (Paris)} \textbf{1927}, \textit{177}, 630--632.
\bibitem{debroglie1924}
De Broglie, L. Sur la d\'efinition g\'en\'erale de la correspondance entre onde et mouvement.  \textit{C.~R. Acad.~Sci. (Paris)} \textbf{1927}, \textit{179}, 39--40.
\bibitem{debroglie1925th}
De Broglie, L. {\it Recherches sur la th\'eorie des quanta}, Facult\'e des Sciences de Paris, 1924; \textit{Annales de Physique} \textbf{1925}, \textit{l0-\`eme s\'erie, III}, 22--128.
\bibitem{debroglie1927}
De Broglie, L. La m\'{e}canique ondulatoire et la structure atomique de la mati\`{e}re et du rayonnement. \textit{J. Phys. Radium} \textbf{1927}, \textit{8}, 225--241;  translated in: De Broglie, L., and Brillouin, L.:  \textit{Selected papers on wave mechanics}, Blackie and Son: Glasgow, UK, 1928.
\bibitem{Valentini}
Bacciagaluppi, G., Valentini, A. \textit{Quantum theory at the crossroads: Reconsidering the 1927 Solvay Conference}, Cambridge University Press: Cambridge, UK, 2009. 
\bibitem{debroglie1930}
De Broglie, L. {\it Introduction \`{a} l'\'{e}tude de la m\'{e}canique ondulatoire}, Hermann: Paris, France, 1930. English translation as \textit{An introduction to the study of wave mechanics}, Methuen: London, UK, 1956. 
\bibitem{Bohm1952}
Bohm, D., A suggested interpretation of the quantum theory in terms of ``Hidden'' Variables. \textit{Phys. Rev.} \textbf{1952}, \textit{85}, 166--179.
\bibitem{Hiley}
Bohm, D., and Hiley, B.~J.:  \textit{The undivided Universe.} Routledge: London, UK, 1993. 
\bibitem{debroglie1956}
De Broglie, L. {\it Une tentative d'interpr\'{e}tation causale et non lin\'{e}aire de la m\'ecanique ondulatoire: la th\'eorie de la double solution}, Gauthier-Villars: Paris, France, 1956;  translated  in {\it Nonlinear wave mechanics: A causal interpretation}, Elsevier: Amsterdam, Netherlands, 1960.
\bibitem{Rosen}
Einstein, A., Rosen, N. The particle problem in the general theory of relativity. \textit{Phys. Rev.}\textbf{1935} \textit{48},
73–-77.
\bibitem{Mie}
Mie, G. Grundlagen einer Theorie der Materie. \textit{Ann. der Phys. (Berlin)} \textbf{1912} \textit{99}, 1--40.
\bibitem{Born}
Born, M., Infeld, L. Foundations of the new field theory. \textit{Proc. R. Soc. A Math. Phys. Eng. Sci.} \textbf{1934}
\textit{144}, 425--451.
\bibitem{Einstein1909}
Einstein, A. \"{U}ber entwicklung unserer anschauungen \"{u}ber das wesen und die konstitution der strahlung. \textit{Physikalische Zeitschrift} \textbf{1909}, \textit{10}, 817--825.
\bibitem{debroglie1925}
De Broglie, L. Sur la fr\'equence propre de l'\'electron.  \textit{C.~R. Acad.~Sci. (Paris)} \textbf{1925}, \textit{180}, 498--500.
\bibitem{debroglie1925b}
De Broglie, L. \textit{Ondes et mouvements}, Gauthier-Villars: Paris, France, 1926. 
\bibitem{Drezet2023}
Drezet, A. A time-symmetric soliton dynamics \`{a} la de Broglie. \textit{Found. Phys.} \textbf{2023}, \textit{53}, 72.
\bibitem{Wheeler}
Wheeler, J. A.  and Feynman, R. P. Interaction with the absorber as the mechanism of radiation. \textit{Rev. Mod. Phys.} \textbf{1945}, \textit{17}, 157--181; Classical electrodynamics in terms of direct interparticle action. \textit{Rev. Mod. Phys} \textbf{1949} \textit{21}, 425--433.
\bibitem{Bell}
Bell, J.S.  \textit{Speakable and Unspeakable in Quantum Mechanics}, 2nd ed.; Cambridge University Press: Cambridge, UK, 2004.
\bibitem{Couder}
Couder, Y., Fort, E. Single-particle diffraction and interference at a macroscopic scale. \textit{Phys. Rev. Lett.} \textbf{2006}, \textit{97}, 154101.`
\bibitem{Bush}
Bush, J.W. The new wave of pilot-wave theory. \textit{Phys. Today} \textbf{2015}, \textit{68}, 47--53.
\bibitem{Jamet1}
Drezet, A., Jamet, P., Bertschy, D., Ralko, A., Poulain, C. Mechanical analog of quantum bradyons
and tachyons. \textit{Phys. Rev. E}  \textbf{2020}, \textit{102}, 052206.
\bibitem{Jamet2}
Jamet, P., Drezet, A.  A mechanical analog of Bohr’s atom based on de Broglie’s double-solution approach. \textit{Chaos} \textbf{2021} \textit{31}, 103120.
\bibitem{Aspect}
Aspect, A.;  Dalibard, J.; Roger, G. Experimental tests of realistic local theories via Bell’s theorem.  \textit{Phys.~Rev.~Lett.} \textbf{1982}, \textit{49}, 1804--1807.
\bibitem{Zeilinger}
Weihs, G.; Jennewein,  T.;  Simon, C.; Weinfurter, H.; Zeilinger, A. Violation of Bell's inequality under strict Einstein locality conditions.  \textit{Phys.~Rev.~Lett.} \textbf{1998}, \textit{81}, 5039--5043.
\bibitem{Hanson}
Hansen B.  et al. Loophole-free Bell inequality violation using electron spins separated by 1.3 kilometres. \textit{Nature} \textbf{2015}, \textit{526}, 682--686.
\bibitem{Gisin}
Salart, D.; Baas, A.; Branciard, C.; Gisin, N.; Zbinden, H. Testing the speed of `spooky action at a distance'. \textit{Nature} \textbf{2008}, \textit{454}, 861--864. 
\bibitem{DrezetFP2019}
Drezet, A.: Lorentz-invariant, retrocausal, and deterministic hidden variables. \textit{Found. Phys.} \textbf{2019}, \textit{49}, 1166--1199. 
\bibitem{Zeilinger2}
D. Rauch et al. Cosmic Bell test using random measurement from high-redshift quasars.  \textit{Phys. Rev. Lett.} \textbf{2018}, \textit{121}, 080403.
\bibitem{Zeilinger3}
Handsteiner, J.  et al. Cosmic Bell test: Measurement settings from milky way  stars. \textit{Phys. Rev. Lett.} \textbf{2017}, \textit{118}, 060401.
\bibitem{Kaiser}
Gallicchio, J.; Friedman, A. S.; Kaiser, D.  Testing Bell's inequality  with comsic photons: Closing the independence loophole.  \textit{Phys. Rev. Lett.} \textbf{2014}, \textit{112}, 110405.
\bibitem{Zeilinger4}
Arndt, M.; Aspelmeyer, M.; Zeilinger, A. How to extend quantum experiments. \textit{Fortschr. Phys.} \textbf{2009}, \textit{57 (11-12)}, 1153--1162.
\bibitem{Hooft}
't Hooft, G. Entangled quantum states  in a local deterministic theory.  arXiv:0908.3408  (2009).
\bibitem{Everett}
Everett, H., III. `Relative State'  formulation of quantum mechanics. \textit{Rev. Mod. Phys.} \textbf{1957}, {\em 29}, 454--462.
\bibitem{Drezetsymmetry} 
Drezet, A. An elementary proof that Everett’s quantum multiverse is nonlocal: Bell-locality and branch-symmetry in the many-worlds interpretation. \textit{Symmetry} \textbf{2023} \textit{15}, 1250.
\bibitem{Vigierthesis}
 Vigier, J.-P. {\it Structure des micro-objets dans l’interpr\'etation causale de la th\'eorie de la th\'eorie des quanta}, Gauthier-Villars: Paris, France, 1956.
\bibitem{Fargue} 
Fargue, D. Permanence of the corpuscular appearance and non linearity of the wave equation.  In: \textit{The wave-particle dualism}; Diner, S.; Fargue, D., Lochak, G., Selleri, F.  Eds.; D. Reidel Publishing: Dordrecht, Netherlands, 1984; 149-172.
\bibitem{DrezetReview} 
Drezet, A. The guidance theorem of de Broglie. \textit{Ann. Fond. de Broglie} \textbf{2021} \textit{46}, 65--85.
\bibitem{dasilva}
Andrade e Silva, J.  \textit{Annales de l'institut Henri Poincar\'e} \textbf{1960} \textit{16}, 289--359.
\bibitem{debroglie1971}
De Broglie, L.  and Andrade e Silva, J.  \textit{La r\'einterprétation de la m\'ecanique ondulatoire. Premi\`ere partie: Principes g\'en\'eraux}, Gauthier-Villars: Paris, France, 1971.
\bibitem{debroglie1974}
De Broglie, L. Sur la r\'efutation du th\'eor\`{e}me de Bell. \textit{C. R. Acad. Sci. (s\'erie B)} \textbf{1974}, \textit{278}, 721--722.
\bibitem{Bellreply}
Bell, J. S.  \textit{Epistemological Letters}, pp. 2--6 (Nov. 1975). Reproduced in \cite{Bell}.
\bibitem{Vigier}
Vigier, J.P. {\it Jean-Pierre Vigier and the
stochastic interpretation of quantum mechanics}; S. Jeffers, B. Lehnert, N. Abramson, L. Chebortarev Eds.; Apeiron: Montreal, Canada, 2000.
\bibitem{Drezet2023b}
Drezet, A. Quantum solitodynamics: Non‐linear wave mechanics and pilot‐wave theory. \textit{Found. Phys.} \textbf{2023}, \textit{53}, 31.
\bibitem{Tetrode}
Tetrode, H. \"Uber den wirkungszusammenhang der welt. Eine erweiterung der klassischen dynamic. \textit{Z. Phys.} \textbf{1922}, \textit{10}, 317--328.
\bibitem{Page}
Page, L. A generalization of electrodynamics with applications to the structure of the electron and to non-radiating orbits. \textit{Phys. Rev.} \textbf{1921},\textit{18}, 292--302; Advanced potentials and their application to atomic models. \textit{Phys. Rev.} \textbf{1924}, \textit{24}, 296--305.
\bibitem{Fokker}
A. D. Fokker, Ein invarianter variationssatz f\"ur die bewegung mehrerer electrischer massenteilshen. \textit{Z. Phys.} \textbf{1929}, \textit{58}, 386--389.
\bibitem{Costa}
Costa de Beauregard, O. A response to the argument directed
by Einstein, Poldosky and Rosen against the Bohrian interpretation
of quantum phenomena, \textit{C. R. Acad. Sci.} \textbf{1953}, \textit{236}, 1632--1634. The idea was developped during the war around 1942-47 but Costa de Beauregard waited for the approval of de Broglie which came after the publication by Wheeler and Feynman~\cite{Wheeler}.
\bibitem{Costa2}
Costa de Beauregard, O. Time symmetry and the Einstein
paradox. \textit{ Nuovo Cimento Soc. Ital. Fis.} \textbf{1977}, \textit{42B}, 41--64; 
 Lorentz and CPT invariances and the Einstein-Podolsky-Rosen correlations. \textit{Phys. Rev. Lett .} \textbf{1983}, \textit{50}, 867--869.
\bibitem{Cramer}
Cramer, J. G.  The transactional interpretation of quantum mechanics. \textit{Rev. Mod. Phys.} \textbf{1986}, 
\textit{58}, 647--687.
\bibitem{Aharonov}
Aharonov, Y.; Gruss, E. Y. Two-time interpretation of quantum mechanics. e-print arXiv:quant-ph/0507269
\bibitem{Sutherland}
Sutherland, R. I. Causally symmetric Bohm model. \textit{Stud. Hist. Philos. Mod. Phys.} \textbf{2008}, \textit{39}, 782--805; Found. Phys. \textbf{47}, 174 (2017).
\bibitem{Sen}
Sen, I.  A local $\psi$-epistemic retrocausal hidden-variable
model of Bell correlations with wavefunctions in physical space. \textit{Found. Phys.} \textbf{2019}, \textit{49}, 83. 
\bibitem{debroglie1927CR}
De Broglie, L. Corpuscules et ondes $\Psi$.  \textit{C.~R. Acad.~Sci. (Paris)} \textbf{1927}, \textit{185}, 1118--1119.
\bibitem{Saari} 
Saari, P. {\it Superluminal localized waves of electromagnetic field in vacuo}, in Mugnai, D., Ranfagni, A., Schulman, L. S. (eds),  {\it Time's Arrows, Quantum Measurement and Superluminal Behavior}, Scientific Monographs: Physics Sciences Series, Italian CNR Publisher: Rome, Italy, 2001, pp. 37--48.
\bibitem{VigierCR}
Vigier, J. P. M\'ecanique ondulatoire dans l'espace de configuration.  \textit{C.~R. Acad.~Sci. (Paris)} \textbf{1952}, \textit{235}, 1372--1375.
\bibitem{RegnierCR}
R\'egnier, A. Sur la conservation de la cahrge.  \textit{C.~R. Acad.~Sci. (Paris)} \textbf{1952}, \textit{235}, 1370--1372.
\bibitem{Durr}
D\"urr, D.; Goldstein, S.; Zangh\`{i}, N. {\it Bohmian mechanics and the meaning of the wave function}, in Cohen, R.S.; Horne, M.; Stachel, J. (eds), {\it Experimental metaphysics—quantum mechanical Studies for Abner Shimony}, Volume One, (Boston Studies in the Philosophy of Science 193), Kluwer Academic Publishers: Boston, USA, 1997.   
\bibitem{DrezetEntropy}
Drezet, A. Justifying Born’s rule $P_\alpha=|\Psi_\alpha|^2$ using deterministic chaos, decoherence, and the de Broglie-Bohm quantum theory. \textit{Entropy} \textbf{2021}, \textit{23}, 1371.
\bibitem{Selleri}
Selleri, F.  \textit{Die debatte um die quantentheorie}, Vieweg+Teubner Verlag Wiesbaden, Springer: 1983.
\bibitem{Hardy}
Hardy, L., Squires, E. On the violation of Lorentz-invariance in deterministic hidden-variable interpretations of quantum mechanics. 
\textit{Phys. Lett. A} \textbf{1992}, \textit{168}, 169-173. 
\bibitem{Palmer}
Palmer, T. Superdeterminism without conspiracy. arXiv:2308.11262 (2023).
\bibitem{Ciepielewski}
Ciepielewski, G.; Okon, E.; Sudarsky, D. On superdeterministic rejections of settings independence. \textit{The British Journal for the Philosophy of Science} \textbf{2023}, \textit{74}, 435–467.
\bibitem{Kyprianidis} 
Kyprianidis, A. Particle trajectories in relativistic quantum mechanics. \textit{Phys. Lett. A} \textbf{1985}, \textit{111}, 111-116.
\end{thebibliography}
\end{document}